\documentclass[twoside,twocolumn, 9pt]{article}

\usepackage[left=1.5cm, right=1.5cm, top=1.785cm, bottom=2.0cm]{geometry}
\usepackage{balance}
\usepackage{times,mathptmx}
\usepackage{graphicx} 

\usepackage[format=plain,justification=justified,singlelinecheck=false,font={stretch=1.125,small,sf},labelfont=bf,labelsep=space]{caption}
\usepackage{float}
\usepackage[english]{babel}
\usepackage{array}
\usepackage[usenames,dvipsnames]{xcolor}
\usepackage{setspace}
\usepackage{amsmath}
\usepackage{amssymb}
\usepackage{siunitx}
\usepackage{textcomp}

\definecolor{cream}{RGB}{222,217,201}
\definecolor{orange}{rgb}{0.8500, 0.3250, 0.0980}
\definecolor{amber}{rgb}{1.0, 0.75, 0.0}
\definecolor{arsenic}{rgb}{0.23, 0.27, 0.29}
\definecolor{battleshipgrey}{rgb}{0.52, 0.52, 0.51}
\definecolor{charcoal}{rgb}{0.21, 0.27, 0.31}
\definecolor{darkelectricblue}{rgb}{0.33, 0.41, 0.47}
\definecolor{firebrick}{rgb}{0.7, 0.13, 0.13}
\definecolor{azure}{rgb}{0.0, 0.5, 1.0}
\definecolor{purple}{rgb}{0.63, 0.36, 0.94}

\begin{document}


\newcommand{\SImum}{\textrm{\textmu{}m}}





\twocolumn[
  \begin{@twocolumnfalse}
\vspace{3cm}

\begin{tabular}{m{0 cm} p{17 cm} }

 & \noindent\LARGE{\textbf{Evaporation-driven liquid flow in sessile droplets}} \\
\vspace{0.3cm} & \vspace{0.3cm} \\

 & \noindent\large{Hanneke Gelderblom\textit{$^{a,d}$}, Christian Diddens\textit{$^{b,d}$} and Alvaro Marin\textit{$^{c,d}$}} \\
\vspace{0.3cm} & \vspace{0.3cm} \\

& \noindent\normalsize{The evaporation of a sessile droplet spontaneously induces an internal capillary liquid flow. The surface-tension driven minimisation of surface area and/or surface-tension differences at the liquid-gas interface caused by evaporation-induced temperature or chemical gradients set the liquid into motion.  This flow drags along suspended material and is one of the keys to control the material deposition in the stain that is left behind by a drying droplet. Applications of this principle range from the control of stain formation in the printing and coating industry, to the analysis of DNA, to forensic and medical research on blood stains, and to the use of evaporation-driven self-assembly for nanotechnology.  Therefore, the evaporation of sessile droplets attracts an enormous interest from not only the fluid dynamics, but also the soft matter, chemistry, biology, engineering, nanotechnology and mathematics communities. As a consequence of this broad interest, knowledge on evaporation-driven flows in drying droplets has remained scattered among the different fields, leading to various misconceptions and misinterpretations. In this review we aim to unify these views, and reflect on the current understanding of evaporation-driven liquid flows in sessile droplets in the light of the most recent experimental and theoretical advances. In addition, we outline open questions and indicate  promising directions for future research.} 

\end{tabular}

 \end{@twocolumnfalse} \vspace{0.6cm}

  ]


\footnotetext{\textit{$^{a}$~Department of Applied Physics and Institute for Complex Molecular Systems, Eindhoven University of Technology, The Netherlands, E-mail: h.gelderblom@tue.nl}}
\footnotetext{\textit{$^{b}$~Physics of Fluids, University of Twente, The Netherlands, E-mail:c.diddens@utwente.nl}}
\footnotetext{\textit{$^{c}$~Physics of Fluids, University of Twente, The Netherlands, E-mail:a.marin@utwente.nl}}
\footnotetext{\textit{$^{d}$~J.M. Burgers Center for Fluid Dynamics, The Netherlands}}







\section{Introduction}

\begin{figure}[h!]
\centering
  \includegraphics[width=0.5\textwidth]{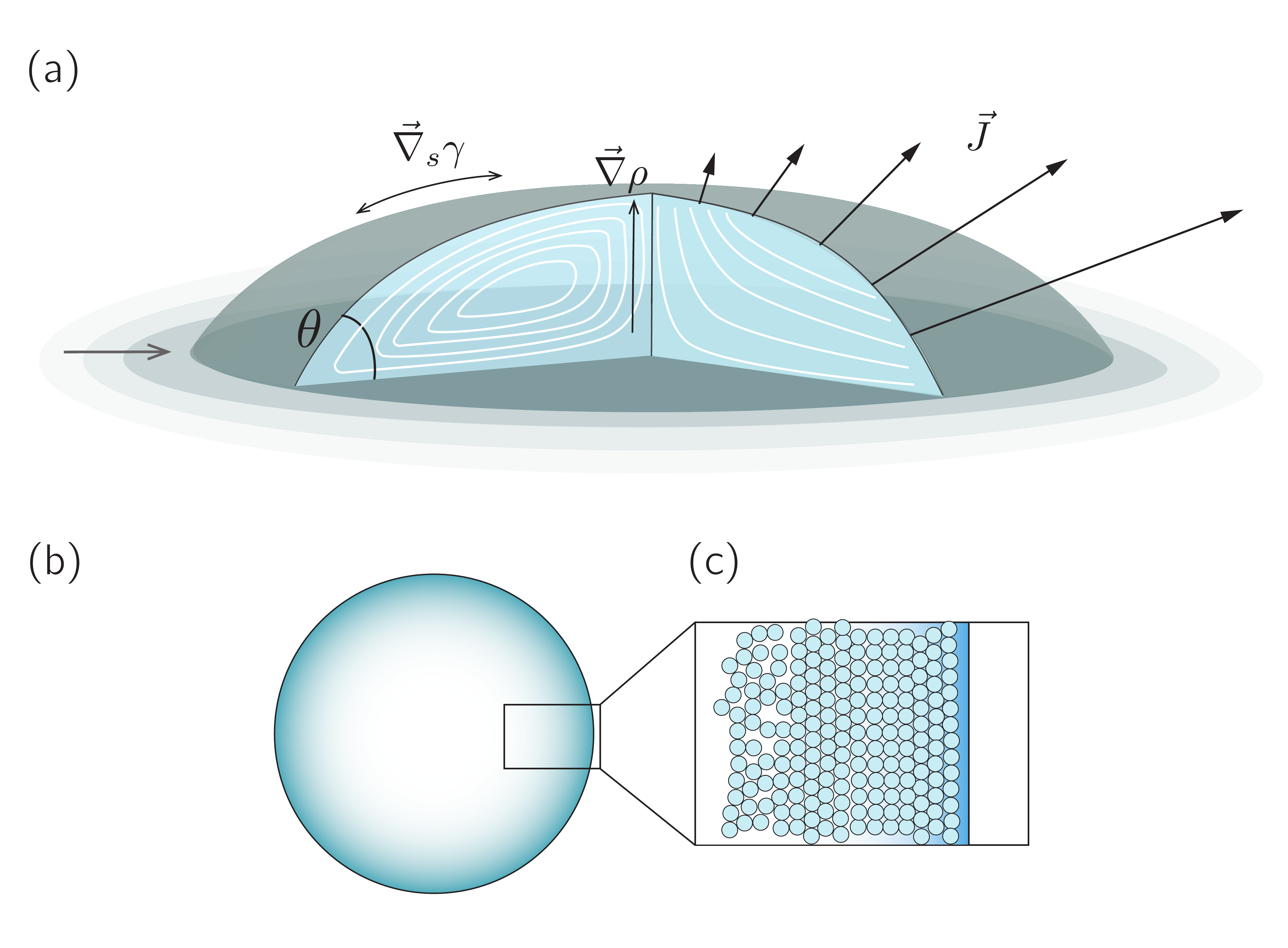}
  \caption{a) Sketch of a generic evaporating sessile droplet with a contact angle $\theta$, and inhomogeneous evaporative flux $\vec{J}$ from its surface. The droplet might experience contact line motion with a velocity $\vec{u}_{cl}$ or/and surface tension gradients along its liquid-air interface $\vec{\nabla_s}\gamma$. In this review we will cover how all these different phenomena influence the velocity field $\vec{u}$ inside the droplet. b) Top view of the droplet showing its circular perimeter. c) Close-up of the contact line. When the droplet contains a diluted suspension of monodisperse colloids and a capillary flow \cite{Deegan:1997} dominates their transport, particles self-organize in a well-defined way \cite{Marin:2011}.}
  \label{fig:introfig}
\end{figure}

The evaporation of a sessile droplet spontaneously induces an internal liquid flow. This flow is the result of several complex phenomena, as illustrated in Figure \ref{fig:introfig}: First, the surface-tension driven minimization of the interfacial area generates a capillary flow to compensate for the evaporative loss from the droplet surface \cite{Deegan:1997}. If the contact line is not pinned, evaporation-induced contact-line motion can couple with the internal flow \cite{Snoeijer:2013}.
Third, the non-uniform evaporative flux could induce temperature and/or solute concentration gradients that in turn give rise to a Marangoni flow \cite{Cammenga:1984}.
Fourth, in liquid mixtures natural convection triggered by evaporation could drive internal flow.\cite{Edwards2018Gravity,Yaxing2019Glycerol} 

The most ubiquitous of these evaporation-driven flows gives rise to the so-called \emph{coffee-stain effect}, which appears when the capillary flow drags suspended particles towards the droplet's contact line. This effect was first demonstrated by Deegan et al.~\cite{Deegan:1997} and opened up an entirely new field of research. Since the pioneering work of the Chicago group\cite{Deegan:1997,deegan2000contact,deegan2000pattern} many studies focused on the evaporation process itself (see reviews \cite{Cazabat:2010, Erbil:2012, Kovalchuk:2014}). Other related works concentrated on wetting and spreading \cite{Bonn:2009} or on contact-line motion \cite{Snoeijer:2013}, with little attention on evaporation. 
A number of studies addressed specifically the evaporation-driven flow inside sessile droplets \cite{Hu:2005p522,Hu2005Marangoni,Hu:2006,Petsi:2008, Masoud:2009, Masoud:2009Stokes, Marin:2011, Marin2011Rush, Gelderblom:2012, Marin2016surfactant}. In his review from 2014, Larson \cite{Larson:2014} showed a selection of the most important contributions to the field from a chemical engineering perspective. 
Indeed, the major boom in the studies on evaporating droplets has occurred in the field of material science and chemical engineering, motivated by the desire to control the shape and structure of the deposits, see e.g. review articles\cite{Sefiane:2010,Han:2012,Sefiane:2014fl,Mampallil:2018} for an overview. The key aspiration in these works is that by controlling the evaporation-driven flow, one could in principle predict and eventually manipulate the distribution of suspended non-volatile material at will. However, due to this practical motivation and the diversity of communities involved, the influence of internal flows in evaporating droplets is often overlooked or misinterpreted, and detailed knowledge remains dispersed among the different fields.

In this review, we discuss the current theoretical, numerical and experimental insights on the flow inside evaporating droplets. We aim to  present a unified view and identify key open questions from a \emph{fluid-dynamic} perspective and discuss its consequences to systems constituted by \textit{soft matter}. We restrict ourselves to the paradigmatic case of droplets evaporating in ambient conditions, where free-convective transport of humid air \cite{Carrier:2016} and evaporative cooling \cite{Dunn:2009} are of negligible influence such that the evaporation is diffusion-limited\cite{Cazabat:2010}, and the droplet shape evolves in a quasi-steady fashion \cite{Gelderblom:2011}. We start from the simplest case of a freely suspended droplet, and then show how the interaction with different types of substrates gives rise to an evaporation-driven capillary flow, and how this flow is influenced by the contact angle and contact-line dynamics. Second, we discuss the influence of the liquid-gas interface on the internal flow, via solutal and thermal Marangoni stress. Third, we address the internal flow due to natural convection in evaporating liquid mixtures. Fourth, we discuss how evaporation-driven flows influence the transport and deposition of a dilute concentration of suspended non-volatile material, e.g. colloidal particles. We close by laying out the main open fluid-dynamical questions and promising directions for future work.

\section{Capillary flow and the role of the substrate} 
Evaporative mass loss from the surface of a drying drop can induce an internal flow, as was first demonstrated in the seminal work by Deegan et al.~\cite{Deegan:1997}. This flow has its origin in capillarity: evaporative mass loss causes the droplet to change its shape, while at the same time it has to maintain its spherical-cap shape dictated by surface tension. As a consequence of the mismatch between these two effects, a capillary flow arises \cite{Deegan:1997}.  However, the presence and nature of such capillary flow depends in a highly non-trivial way on the evaporative flux profile\cite{Fischer:2002, Masoud:2009} and the interaction of the droplet with a substrate, i.e.~on the contact angle\cite{Petsi:2008,Masoud:2009,Gelderblom:2012} and the motion of the contact line\cite{Petsi:2008,Berteloot:2008,Masoud:2009}. In the literature, this complexity often leads to confusion when addressing the flow direction (inward, outward, or circulatory) inside an evaporating droplet.

In this section we will discuss the influence of the liquid-substrate interaction on the capillary flow step by step, always assuming thermal equilibrium among all phases involved. We start from the simplest situation: a freely suspended evaporating droplet. Then, we move to the situation where the droplet is in contact with a substrate.  We separately discuss the influence of the contact angle, pinning and motion of the contact line on the capillary flow, and argue that -perhaps counter-intuitively- neither contact-line pinning nor a diverging evaporative flux at the contact line are essential for the existence of a radially outward capillary flow. On partially wetting substrates, two situations will be analyzed in detail: the freely moving contact line, in which the contact angle remains constant (CCA or constant contact-angle mode), and the pinned contact line, in which the wetted area remains constant (CCR or constant contact-radius mode)\cite{Picknett:1977p108, Stauber:2015}. Finally, we discuss the influence of complete wetting and hydrophobic substrates on the capillary flow.


\subsection{Freely suspended droplet}

\begin{figure}[h!]
\centering
  \includegraphics[width=0.25\textwidth]{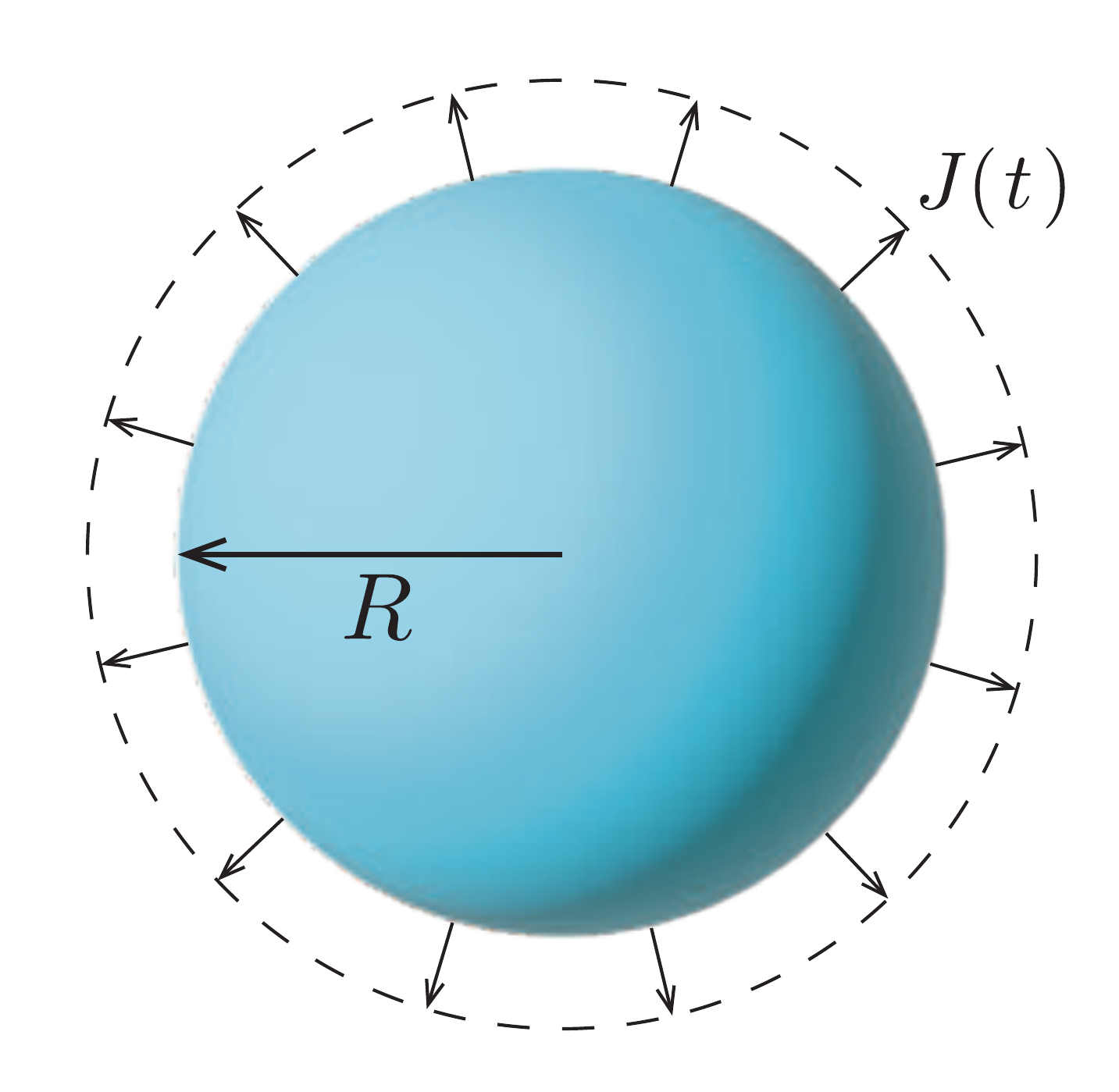}
  \caption{Uniform evaporation from the surface of a freely suspended droplet causes the inward motion of the liquid-air interface.
  }
  \label{fig:spherical}
\end{figure}

The simplest geometry to consider is a droplet that is freely suspended in quiescent air, as illustrated in Fig.\,\ref{fig:spherical}. This configuration, which has first been studied by Maxwell \cite{maxwell:1877} and Langmuir \cite{langmuir:1918}, is found in e.g.~spray drying applications, aerosols, airborne disease transmission, combustion and atmospheric science (see e.g. \cite{Sen:2009,Cazabat:2010, Sadek:2015, Ghabache:2016, Bourouiba:2021} and references therein). In absence of a substrate, the evaporative flux from the droplet surface is uniform and radially outward. In the diffusion-limited regime the evaporative flux is given by \cite{Cazabat:2010}
\begin{equation}
    J=\frac{D \Delta c}{R},\label{uniflux}
\end{equation}
with $D$ the diffusion coefficient of vapor in air, $R$ the droplet radius, and $\Delta c=c_{s}-c_\infty$ the difference between the saturated vapour concentration just above the liquid-air interface and the ambient vapour concentration far from the droplet. 
The rate of mass loss from the droplet follows by multiplication of the flux with the droplet surface area, from which we find
\begin{equation}
\frac{dM}{dt}=-4\pi R D\Delta c.
\label{eq:dmdt}
\end{equation}
Hence, we observe that since the flux scales as $J\sim R^{-1}$, the rate of mass loss is \emph{not} proportional to the surface area of the drop but to its radius, in contrast to what is often naively anticipated.

Upon combining (\ref{eq:dmdt}) with the expression for the shape change $dM/dt=4\pi \rho R^2 dR/dt$, where $\rho$ the density of the liquid, we find that $dR/dt=-J/\rho$. Upon integration we arrive at the famous $R^2-$law
\begin{equation}
    R^2(t)=R_0^2-2\frac{D\Delta c}{\rho}t
\end{equation}
for a freely suspended evaporating droplet.

Importantly, the evaporative flux from the surface of a freely suspended drop does \emph{not} induce any flow inside the droplet. As the evaporation proceeds, the interface of the droplet is moving inwards to accommodate for the mass loss. One could think about this situation as that the evaporation is just ``peeling off a layer of liquid" from the droplet. Hence, the particle agglomeration at the interface of a colloidal suspension droplet in spray drying conditions (see e.g.~\cite{Pauchard:2004, Sen:2007, Sen:2009, Tsapis:2002, Tsapis:2005, Boulogne:2013, Loussert:2016}) is purely due to the inward motion of the liquid-air interface that collects the particles and not to an evaporation-induced internal flow, as is sometimes suggested in the literature. Indeed, it is not the particles that get advected to the interface, but vice versa: the interface moves inwards, carrying the particles along.

\subsection{Sessile droplet with a contact angle $\theta=90^\circ$}
We now turn to the situation where the droplet is in contact with a substrate: a sessile droplet. 
For illustrative purposes, the configuration of a sessile droplet with a contact angle $\theta=90^\circ$  is described first. In that case, the evaporative flux from the surface is still uniform and given by (\ref{uniflux}), as mathematically, the impermeable substrate can be considered as a mirroring surface \cite{Deegan:1997}. The velocity field inside an evaporating droplet in CCA and CCR mode subject to a uniform evaporative flux has been studied both under the assumption of a potential flow\cite{Petsi:2005, Tarasevich:2005, Petsi:2006, Masoud:2009} and in Stokes flow \cite{Petsi:2008}. In these studies the general evaporation problem is simplified dramatically by assuming a uniform evaporative flux, however, in case $\theta=90^\circ$ the Stokes flow solution\cite{Petsi:2008} becomes exact. 

\subsubsection{Freely moving contact line (CCA mode)}
\begin{figure}[ht]
\centering
  \includegraphics[width=0.45\textwidth]{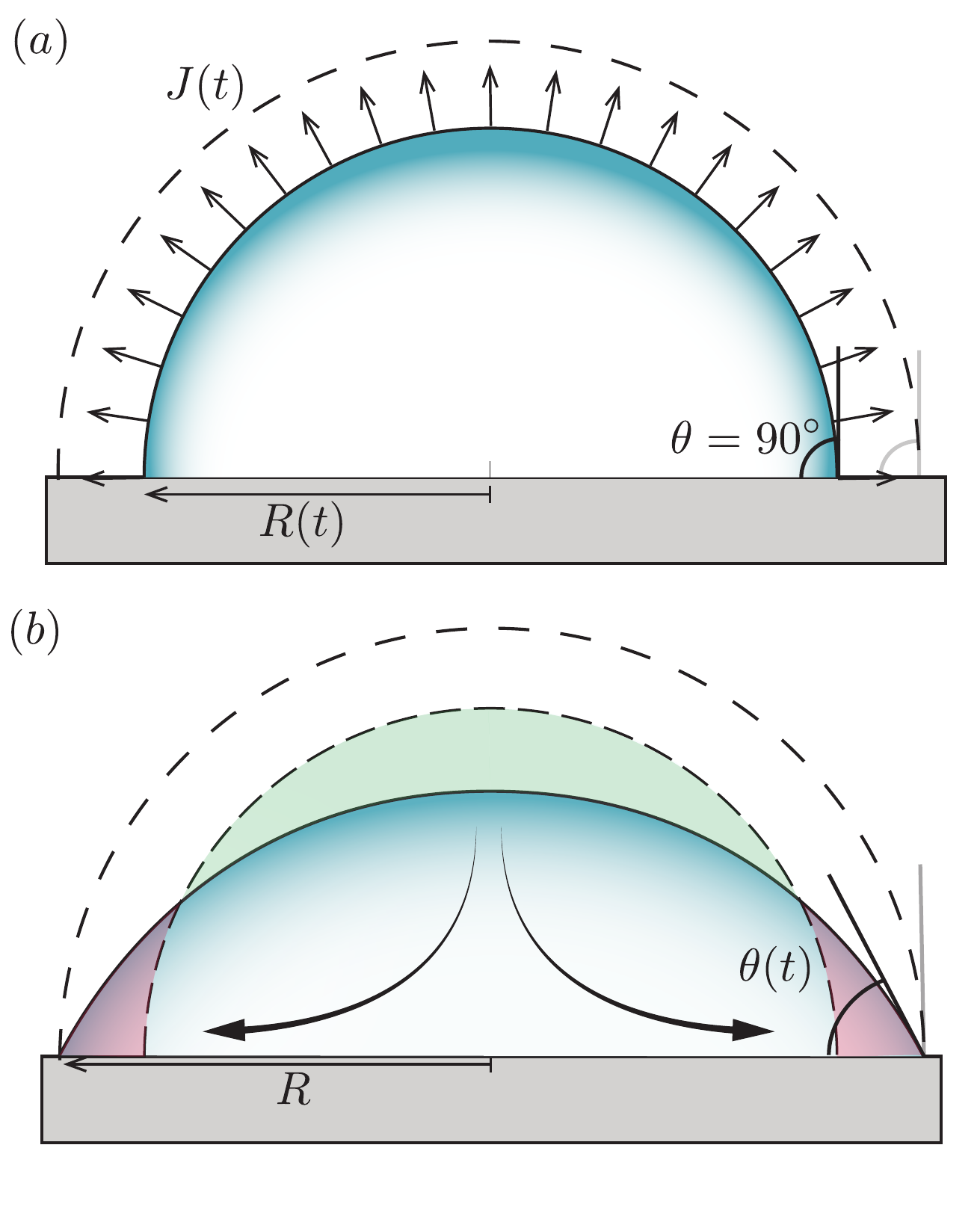}
  \caption{Uniform evaporation from a sessile droplet with a contact angle of $90^\circ$. (a) When the contact line freely recedes the droplet evaporates with a constant contact-angle (CCA), and the interface moves from the dashed to the solid contour. (b) When the contact line is pinned, the droplet evaporates with a constant contact radius (CCR, solid line). According to the uniform evaporative flux, the droplet interface should recede from the dashed contour to the dashed-dotted one (similar to panel a). However, this motion is prevented by the contact-line pinning. To maintain a spherical-cap shaped droplet as dictated by surface tension, a compensatory capillary flow is generated transporting the excess liquid on the top (green area) to replenish the evaporated liquid at the contact line (red area), as illustrated by the arrows.}
  \label{fig:sessile}
\end{figure}

If the contact line can freely recede (dashed contour in Fig.~\ref{fig:sessile}(a)) at a speed that exactly matches the evaporative loss from the surface (i.e.~$dR/dt=-J/\rho$) the drop is evaporating with a constant contact angle, i.e.~in CCA mode. This situation is exactly the same as for the freely suspended drop: there is no capillary mechanism at play to generate a flow and the interface just recedes as dictated by the evaporative flux \cite{Petsi:2008, Masoud:2009}.  

\subsubsection{Pinned contact line (CCR mode)}
When the contact line is pinned (Fig.~\ref{fig:sessile}(b)) the picture drastically changes. Clearly, peeling off a uniform layer of liquid from the droplet surface, as dictated by the evaporative flux, is incompatible with the constraint of a pinned contact line. As illustrated in Fig.~\ref{fig:sessile}(b) a compensatory internal flow\cite{Deegan:1997} is generated to transport the excess liquid from the top of the droplet (green area) to the contact line area (red area). This mechanism is at the heart of the capillary flow inside evaporating sessile droplets. 

Figure \ref{fig:sessile} illustrates that, in contrast to what is often believed, a divergence of the evaporative flux at the contact line is \emph{not} required for the generation of a capillary flow. Indeed, outward flows are also found when the evaporative flux is uniform \cite{Petsi:2005, Tarasevich:2005, Petsi:2006, Petsi:2008}. Whenever there is a mismatch between the local evaporation rate and the constraints posed by the droplet's spherical-cap shape and contact-line motion, a capillary flow arises. Clearly, as soon as the droplet starts to loose mass the contact angle will decrease below $90^\circ$, which is the situation that will be considered next.

\subsection{Sessile droplet on a partially wetting substrate ($0<\theta< 90^\circ$)}\label{sec:partialwetting}
\begin{figure}[ht]
\centering
  \includegraphics[width=0.5\textwidth]{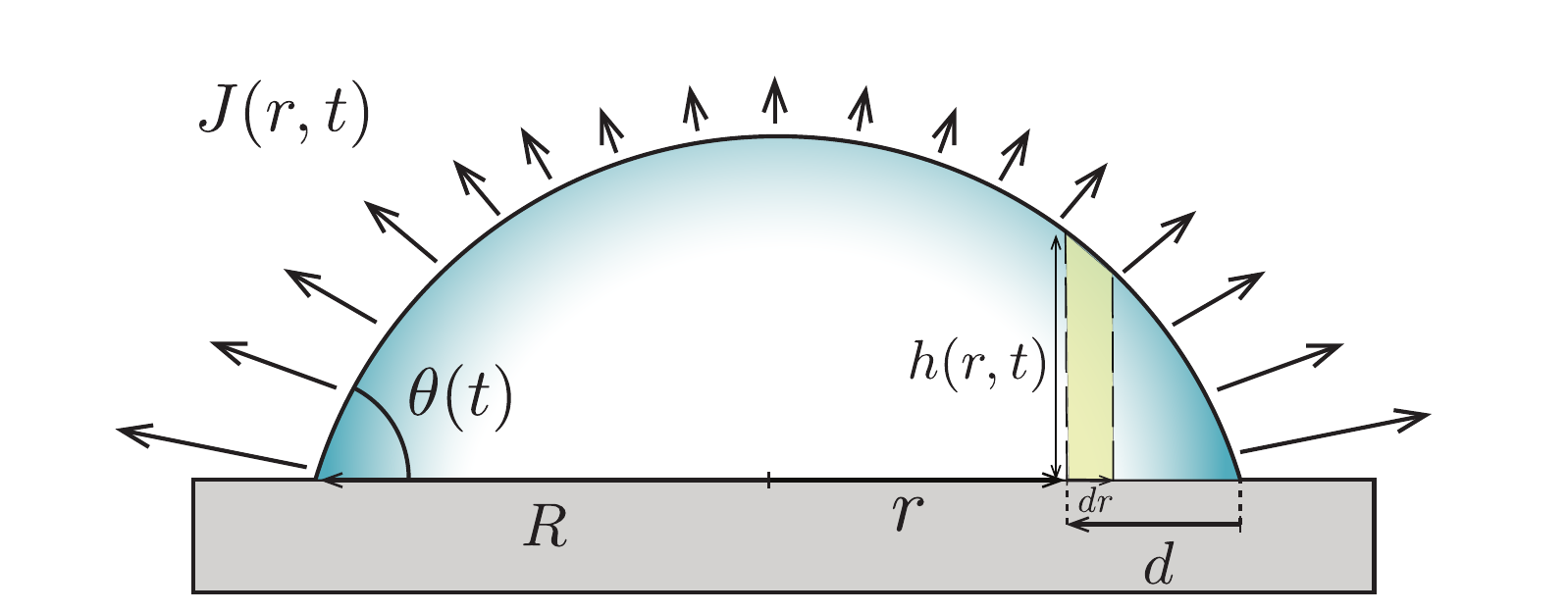}
  \caption{A droplet with base radius $R$ evaporates with a pinned contact line from a partially wetting substrate ($\theta<90^\circ$). The evaporative flux $J$ drives a capillary flow $Q$ inside the droplet with height profile $h(r,t)$. The dashed lines mark a control volume of width $dr$ at a small distance from the contact line $d$.}
  \label{fig:class}
\end{figure}

In the classical example of a pinned droplet on a partially wetting substrate  (finite contact angle $\theta< 90^\circ$, Fig.~\ref{fig:class}), the singular corner geometry of the droplet leads to a divergence of the evaporative flux at the contact line  \cite{Deegan:1997, Bocquet:2007} and the expression (\ref{uniflux}) does not hold anymore. 

We will consider here the case where $\theta\ll 90^\circ$, such that the drop shape is relatively shallow (i.e.~its height is much smaller than its base radius, $h\ll R$). In this situation, the capillary flow inside the droplet can be calculated analytically \cite{Hu:2005p522, Marin:2011}. For this exemplary case, we will summarize the calculation here, considering both the CCR and the CCA modes. 

Consider the axisymmetric, shallow droplet sketched in Fig.~\ref{fig:class}. Following \cite{deegan2000contact, Marin:2011} we define an infinitesimally small control volume of width $dr$ at a distance $r$ from the droplet center. Mass conservation requires that the rate of change of the amount of liquid inside this control volume is equal to the net inflow of liquid minus the amount of liquid that evaporates from the droplet surface:
\begin{equation}
    \frac{\partial h}{\partial t}=-\frac{1}{r}\frac{\partial Q}{\partial r}-\frac{1}{\rho}J, \label{eq:masscons}
\end{equation}
where $h(r,t)$ is the droplet height profile, $t$ the time, and $Q(r,t)$ the volume flow. From (\ref{eq:masscons}) one immediately observes that when the local decrease in droplet height is smaller than the evaporative loss, a flow arises. 

\begin{figure}[ht]
  \includegraphics[trim = 0 0 -15 0, width=0.5\textwidth]{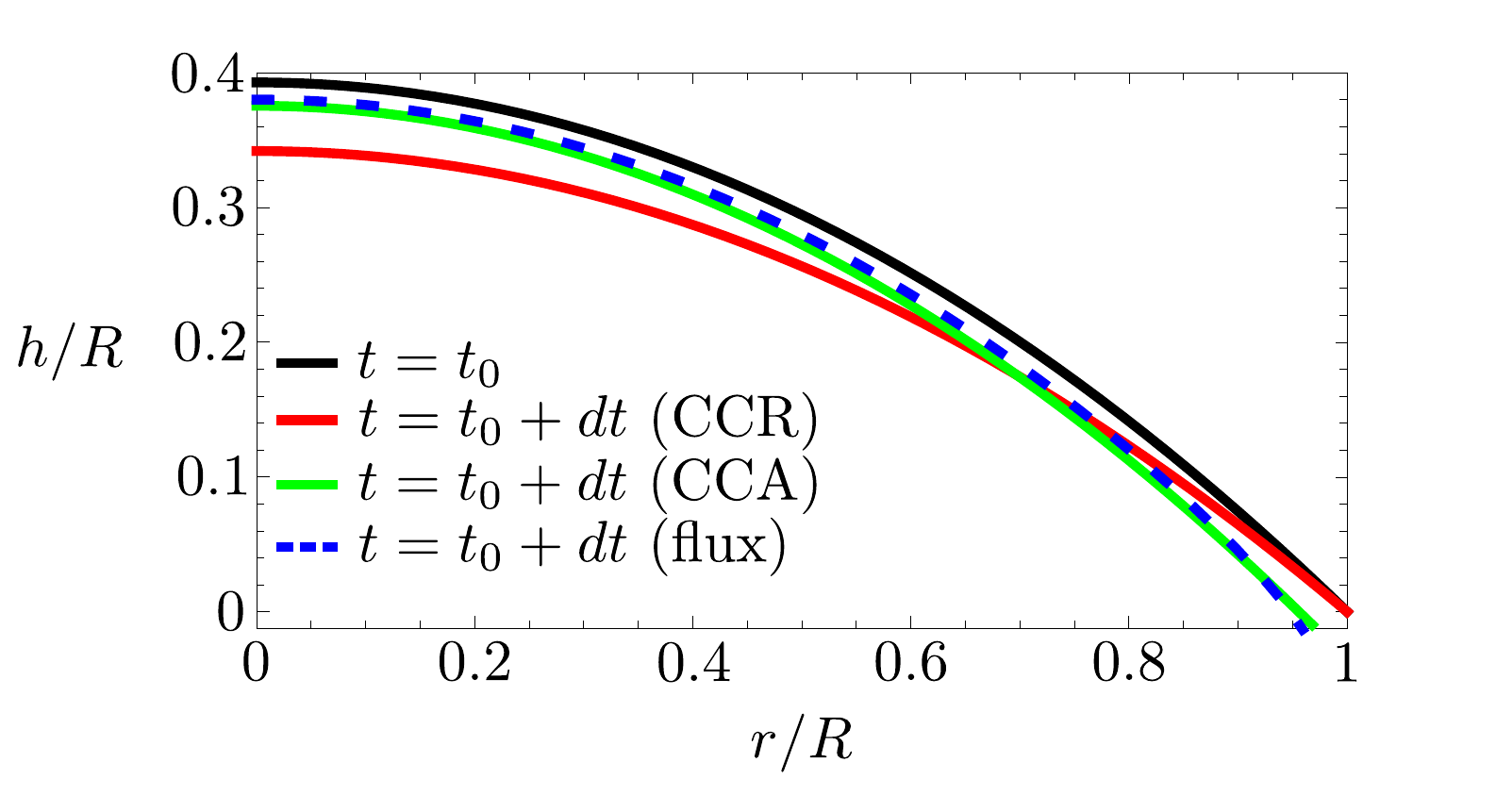}
  \caption{Right half of an evaporating droplet with an initial contact angle of $45^\circ$. Black solid line: droplet contour at $t=t_0$. Red solid line: location of the contour for a pinned contact line a small timestep $dt/\tau=0.02$, where $\tau=D\Delta c/\rho R^2$, later. Green solid line: contour position for a freely moving contact line. Blue dashed line: imaginary contour position imposed by the evaporative flux profile. Note that both the red (CCR mode) and the green (CCA mode) line show a mismatch with respect to the evaporative flux profile in blue. As a consequence, in both CCA and CCR mode a capillary flow will arise.}
  \label{fig:mismatch}
\end{figure}

In the following, we will consider the situation where the droplet shape is not affected by the flow inside the droplet or by gravity, which means that both the capillary number $\mu U/\gamma$ and the Bond number $\rho g R^2/\gamma$, where $\gamma$ is the surface tension, $\mu$ the dynamic viscosity, $U$ the typical velocity scale and $g$ the gravitation acceleration, are small \cite{Larson:2014}. Hence, the droplet shape is restricted to a spherical cap as dictated by surface tension, with either a pinned or a moving contact line (i.e.~evaporation in CCA or CCR mode). Consequently, the local change in droplet height (left-hand side of (\ref{eq:masscons})) is also restricted. Whenever this local height change does not match the evaporative flux a capillary flow arises, as illustrated in Fig.~\ref{fig:mismatch}. 

In the limit of small contact angles ($\theta\ \ll 90^\circ$) simplified expressions for $h$ and $J$ are available, which allows for an explicit analytical solution to (\ref{eq:masscons}). For $\theta\ \ll 90^\circ$ the evaporative flux is given by \cite{deegan2000contact, Marin:2011}
\begin{equation}
    J(r)=\frac{2}{\pi}\frac{D \Delta c}{\sqrt{R^2-r^2}},\label{eq:flux1}
\end{equation}
with $r$ the radial coordinate and $R$ the base radius of the sessile droplet (see Fig.~\ref{fig:class}). For small contact angles the droplet's spherical-cap shape is well described by a parabola with contact angle $\theta$
\begin{equation}
    h(r,t)=\frac{R^2(t)-r^2}{2R(t)}\theta(t)\label{eq:h}
\end{equation}

To obtain an expression for the capillary flow $Q$ using (\ref{eq:masscons}-\ref{eq:h}), we now discriminate again between the case of a pinned and a moving contact line.

\subsubsection{Pinned contact line (CCR mode)}\label{sec:pinned}
When the contact line is pinned, the droplet's base radius $R$ is constant while its contact angle $\theta(t)$ decreases with time. We now use (\ref{eq:masscons}) to calculate the flow and velocity field in this case following the approach detailed in \cite{Marin:2011}. An expression for $d\theta/dt$ and hence $\partial h/\partial t$ in (\ref{eq:masscons}) follows from a global mass conservation: the decrease in droplet volume
\begin{equation}
    \frac{dV}{dt}=\frac{d}{dt}\int_0^Rh(r,t)2\pi r\mathrm{d}r=\frac{\pi R^3}{4}\frac{d\theta}{dt},\label{eq:rtheta}
\end{equation}
equals the total amount of evaporated liquid
\begin{equation}
    \frac{dV}{dt}=-\frac{1}{\rho}\int_0^RJ(r)2\pi r \mathrm{d}r=-\frac{4RD\Delta c}{\rho},\label{eq:massloss}
\end{equation}
such that 
\begin{equation}
   \frac{d\theta}{dt}=-\frac{16D\Delta c}{\pi R^2\rho}\label{eq:dtheta},
\end{equation}
and hence 
\begin{equation}
    \theta(t)=\frac{16D\Delta c}{\pi R^2\rho}(t_f-t),\label{eq:theta}
\end{equation}
with $t_f$ the total lifetime of the droplet. By integration of (\ref{eq:masscons}) and using (\ref{eq:flux1}), (\ref{eq:h}), and (\ref{eq:dtheta}) we find that the radially outward flow is constant in time and given by
\begin{equation}
    Q(r)=\frac{2RD\Delta c}{\pi\rho}\left[\sqrt{1-(r/R)^2}-\left(1-(r/R)^2\right)^2\right].\label{eq:flow}
\end{equation}
The height-averaged radial velocity then follows from 
\begin{equation}
    \overline{u}(r,t)=\frac{Q}{rh}=\frac{4D\Delta c}{\pi\rho r}\frac{1}{\theta(t)}\left[\frac{1}{\sqrt{1-(r/R)^2}}-\left(1-(r/R)^2\right)\right].\label{eq:avu}
\end{equation}
Close to the contact line ($r\to R$) this expression for the height-averaged velocity reduces to
\begin{equation}
    \overline{u}_{\mathrm{app}}(r,t)=\frac{2\sqrt{2}D\Delta c}{\pi\rho R}\frac{1}{\theta(t)}\frac{1}{\sqrt{1-r/R}}.\label{eq:avuapp}
\end{equation}
Once the height-averaged velocity is known, the velocity field within the droplet can be obtained in the lubrication approximation \cite{Hu:2005p522,Gelderblom:2012}.
In that case the Navier-Stokes equations reduce to\cite{Oron:1997}
\begin{equation}
    \frac{dp}{dr}=\mu \frac{\partial^2 u}{\partial z^2},\label{eq:lub}
\end{equation}
with $p$ the pressure, $u$ the radial velocity and $z$ the coordinate perpendicular to the substrate. As boundary conditions one imposes $u(r,0)=0$ (no slip) and $\left.\tfrac{\partial u}{\partial z}\right|_{z=h(r,t)}=0$ (no shear stress at the liquid-air interface). Upon integration of (\ref{eq:lub}) we then find
\begin{equation}
   u(r,z,t)=\frac{3}{h^2(r,t)}\overline{u}(r,t)\left(h(r,t)z-\frac{1}{2}z^2\right),\label{eq:vel} 
\end{equation}
with $\overline{u}$ given by (\ref{eq:avu}). 
An analysis of the full Stokes-flow problem in the corner geometry near the contact line \cite{Gelderblom:2012} has demonstrated that the lubrication velocity field given by (\ref{eq:vel}) is accurate all the way down to the contact line as long as the contact angle of the droplet $\theta\ll 90^\circ$. Moreover, this velocity field is found to be in excellent agreement with Particle Image Velocimetry measurements \cite{Marin:2011,bodiguel2013imaging}. 

The velocity field (\ref{eq:vel}) displays two singularities \cite{Marin:2011}: a \emph{spatial} singularity that originates from the divergence of the evaporative flux (\ref{eq:flux1}) and a \emph{temporal} singularity that originates from the vanishing droplet height towards the end of the droplet lifetime. This temporal singularity has dramatic consequences for the particle deposition in the ring stain that forms\cite{Marin:2011}, as will be discussed in \S \ref{sec:particles}.

\subsubsection{Freely moving contact line (CCA mode)}\label{sec:cca}
A similar calculation can be done for a droplet in CCA mode. In that case we obtain from (\ref{eq:rtheta}) $dV/dt=(3/4)\pi \theta R^2dR/dt$ and hence, through (\ref{eq:massloss}) 
\begin{equation}
    \frac{dR}{dt}=-\frac{16D\Delta c}{3\pi \theta R\rho},\label{eq:dRdt}
\end{equation}
from which
\begin{equation}
  R(t)=\sqrt{R_0^2-\frac{32D\Delta c}{3\pi \theta \rho}t},  
\end{equation}
with $R_0$ the initial droplet radius. Combining (\ref{eq:flux1}), (\ref{eq:h}) and (\ref{eq:dRdt}) we obtain from (\ref{eq:masscons}) by integration



\begin{equation}\label{eq:qmov}
\begin{split}
Q(r,t) &= \frac{2R(t)D\Delta c}{\pi\rho} \left[\sqrt{1-\left(\frac{r}{R(t)}\right)^2}-1 +\right.\\
         &\left.+ \frac{2}{3}\left(\frac{r}{R(t)}\right)^2+\frac{1}{3}\left(\frac{r}{R(t)}\right)^4\right].
\end{split}
\end{equation}


The velocity field is again given by (\ref{eq:vel}), but now with
the height-averaged velocity expressed as
    \begin{equation}\label{eq:avumove}
    \begin{split}
    \overline{u}(r,t)&=\frac{4D\Delta c}{\pi\rho \theta r }\left[\frac{1}{\sqrt{1-(r/R(t))^2}}+\right.\\
    &\left.+\frac{\frac{2}{3}\left(r/R(t)\right)^2+\frac{1}{3}\left(r/R(t)\right)^4-1}{1-(r/R(t))^2}\right].
    \end{split}
\end{equation}

\begin{figure}[ht]
\centering
  \includegraphics[width=0.45\textwidth]{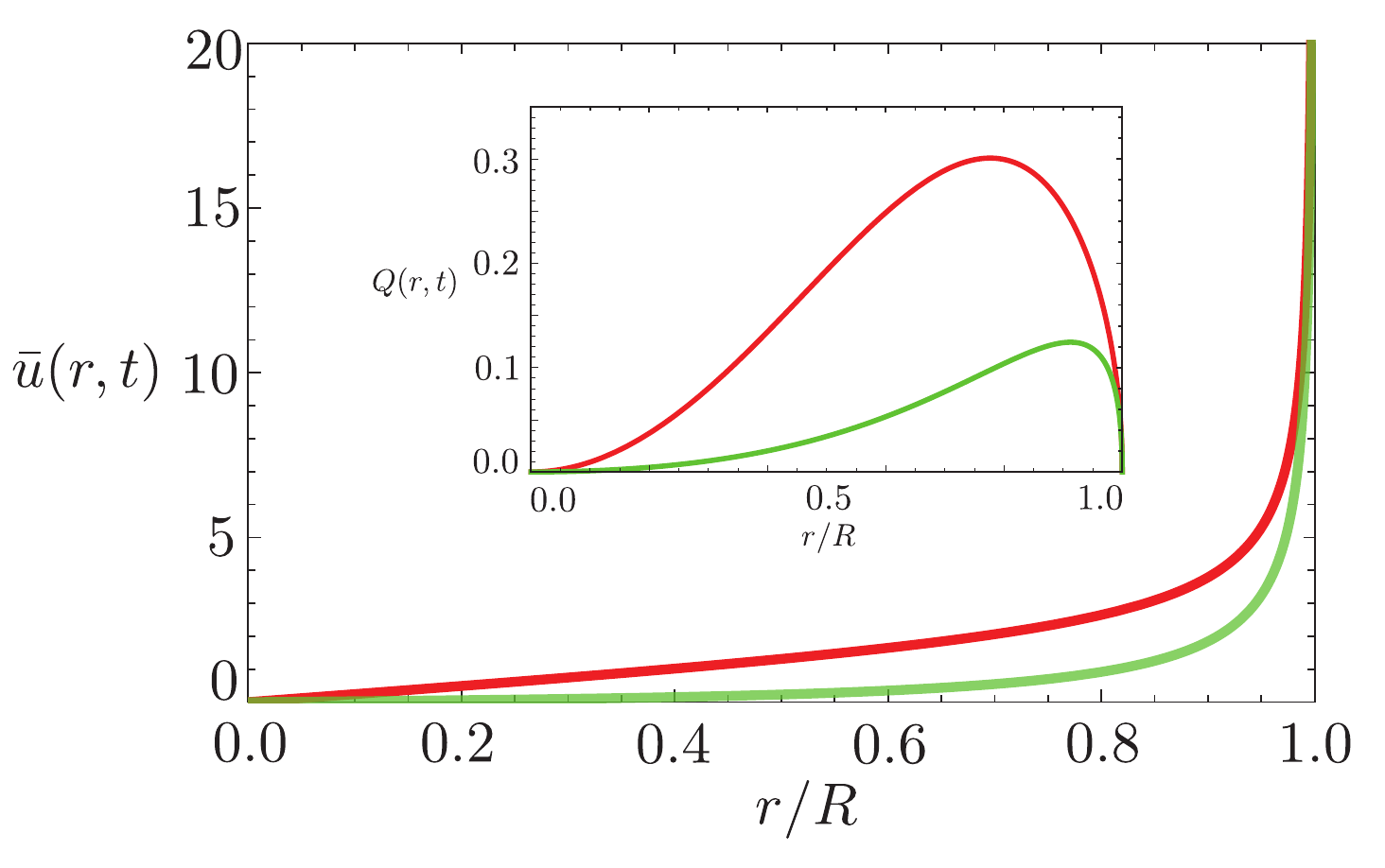}
  \caption{Height-averaged radial velocity as a function of the radial position in a spherical-cap shaped droplet with an (initial) contact angle of $\theta= 45^\circ$. Red solid line: pinned contact line, green solid line: moving contact line. Note that both velocities are directed radially outward. Inset: Radial volume flow as a function of the radial position in the droplet for a pinned (red) and moving (green) contact line. Note that both volume flows are pointing radially outward.}
  \label{fig:avvel}
\end{figure}

In Fig.\,\ref{fig:avvel} we compare the height-averaged velocities for a droplet in CCR mode (\ref{eq:avu}) and CCA mode (\ref{eq:avumove}). In the inset, the same is done for the flows in CCR (\ref{eq:flow}) and CCA mode (\ref{eq:qmov}). Clearly, in both cases (CCR and CCA) the flow and velocity are directed radially outward, as was already expected from the sketch in Fig.\,\ref{fig:mismatch}. Hence, even though in CCA mode the \emph{contact-line itself} is receding, the internal flow is directed radially outward. However, the magnitude of the flow and radial velocity are much larger in CCR mode, as a result of the larger mismatch between the evaporative flux profile and the constrained motion of the liquid-air interface (compare the blue dashed with the green and red solid lines in Fig.\,\ref{fig:mismatch}). 


\subsubsection{Some remarks on the spherical-cap constraint and contact-line motion}
In the analysis above we assumed that surface tension is so strong that the drop maintains a spherical-cap shape independent of the internal flow (i.e.\ the capillary number is small\cite{Snoeijer:2013}) and that, in CCA mode, the contact line is completely free to move. However, these assumptions are not always justified. For a detailed discussion on the coupling between the interface shape, contact-line motion and the flow inside a droplet the reader is referred to the review articles\cite{Bonn:2009,Snoeijer:2013}; here we restrict ourselves to a few comments specific to evaporating droplets. 

First of all, the contact-line motion is often not completely free, but restricted by properties of the substrate (i.e.~the substrate's receding angle)\cite{Stauber:2015} and potential self-pinning due to the deposition of solutes\cite{deegan2000pattern}. Pinning forces could therefore induce a stick-slide/stick-slip motion of the contact line \cite{Berteloot:2008,Stauber:2015}. During the sliding phase there will be an inward flow away from the contact line, as described by Huh \& Scriven\cite{Huh:1971}. At the same time, the evaporation flux from the droplet surface causes an outward flow, as discussed above. Since both flows are governed by the Stokes equations, the two solutions can be superimposed\cite{Gelderblom:2012}, and one would expect a cross-over length scale where the two velocity fields cancel each other\cite{Berteloot:2008, Snoeijer:2013}. Using the previously derived expression for the capillary flow close to the contact line (\ref{eq:avuapp}), we find for this cross-over length scale\cite{Berteloot:2008}
\begin{equation}
    \frac{\ell_c}{R_0}\sim\left(\frac{D\Delta c}{\rho \theta R_0  V}\right)^2,
\end{equation}
where $V$ is the contact-line speed. Beyond this length scale, the flow is dominated by the receding motion of the contact line and directed inwards, while closer to the contact line the evaporation-driven outward flow dominates.
Note that when the contact-line can move freely and its velocity $V$ is given by (\ref{eq:dRdt}), a cross-over no longer occurs and the entire flow field is pointing radially outward, as shown in \S\ref{sec:cca}.

Second, for droplets evaporating in CCA mode, a deformation of the interface will occur close to the contact line: the no-slip boundary condition at the solid substrate conflicts with the moving contact-line condition, causing a divergence of the internal pressure \cite{Huh:1971} and hence a local deformation of the interface. This interface deformation occurs only close to the contact line, while the remainder of the droplet is still described by a spherical cap receding with a macroscopic or apparent contact angle close to its equilibrium value \cite{Bonn:2009}. However, when the equilibrium contact angle is small ($\theta< 5^\circ$)\cite{Berteloot:2008}, the interfacial deformation becomes more prominent and the macroscopic contact angle changes, such that expressions (\ref{eq:qmov}) and (\ref{eq:avumove}) no longer apply. 

Third, also the incompatibility of the diverging evaporative flux with the no-slip condition at the boundary causes a diverging internal pressure field \cite{Gelderblom:2012}. This condition applies to droplets in both CCA and CCR mode, and may again cause interface deformations. However, again the effect is only prominent for small contact angles ($\theta< 5^\circ$)\cite{Berteloot:2008} and at small distances ($\ll R$) from the contact line\cite{Gelderblom:2012}. We note that, importantly, the introduction of a slip length will not cure this pressure divergence as it is even stronger than the one found in the  moving contact line problem\cite{Gelderblom:2012}.

\subsection{Complete wetting ($\theta=0^\circ$)}
When the droplet is completely wetting the substrate, the influence of evaporation-induced interface deformations increases dramatically\cite{Berteloot:2008}. In this case, the equilibrium contact angle of the droplet is zero and will never be reached: while the droplet tries to spread out, evaporation causes the droplet's apparent contact angle to remain finite\cite{Eggers:2010}. Moreover, the droplet shape is no longer constrained to a spherical cap, but evolves over time as a function of both the evaporative flux and internal flow. In turn, this change in drop shape affects the evaporative-flux profile and internal flow \cite{Eggers:2010, Morris:2014}. In addition, as the droplet spreads out to a thin film, disjoining pressure will start to play a role\cite{Cazabat:2010}. As a result the shape of the droplet is not known a priori\cite{Bonn:2009}, which poses a challenging problem: the droplet shape, internal flow and evaporative flux are coupled, and local solutions close to the contact line\cite{Pham:2010, Guena:2007} depend on the global evaporation characteristics\cite{Eggers:2010}. An approximate description of the spreading dynamics based on a pre-described evaporative-flux profile combined with a phenomenological description of the time-dependent droplet radius has been derived by \cite{Poulard:2003, Poulard:2005, Poulard:2005b, Guena:2007}.
A self-consistent theoretical description of the complex coupling that arises between the droplet shape, internal flow and evaporative flux in complete wetting is provided by\cite{Eggers:2010, Morris:2014}. However, in complete wetting, the shape of the droplet, its apparent contact angle, the spreading dynamics and the width of the contact line region still remain subject of debate\cite{Morris:2014,Eggers:2010, Guena:2007, Pham:2010}.

\subsection{Sessile droplet on a hydrophobic substrate ($\theta>90^\circ$)}
For arbitrary contact angles, the evaporative flux in the wedge-shaped region close to the contact line ($(1-r/R)\ll 1$) can be approximated as \cite{Deegan:1997, Gelderblom:2012}
\begin{equation}
    J(r)=A(\theta) \frac{D\Delta c}{ R}\left(1-r/R\right)^{\lambda(\theta)},\label{eq:flux2}
\end{equation}
with $\lambda(\theta)=\pi/\left(2\pi-2\theta\right)$, and $A(\theta)$ a prefactor of order unity that can be obtained from matching to the full spherical-cap solution for the evaporative flux, as derived by Popov\cite{Popov:2005}.
From (\ref{eq:flux2}) one observes that for $\theta>\SI{90}{\degree}$ the evaporative flux no longer diverges (as it does for $\theta<90^\circ$), but decays to zero at the contact line. 

For large contact angles, the calculation of the internal flow becomes extremely complicated. First of all, the description of the evaporative flux and droplet shape are difficult and require the use of toroidal coordinates. An analytical solution for the evaporative flux along the entire spherical-cap surface has been derived by Popov\cite{Popov:2005} based on the solution for the electro-static potential of a lens-shaped, charged conductor provided by Lebedev\cite{Lebedev}.  Second, the mass conservation equation (\ref{eq:masscons}), in which one implicitly assumes $h\ll R$, is no longer applicable. As a consequence, the description of the internal flow becomes challenging and exact explicit solutions are not available.

\begin{figure}[ht]
\centering
  \includegraphics[width=0.4\textwidth]{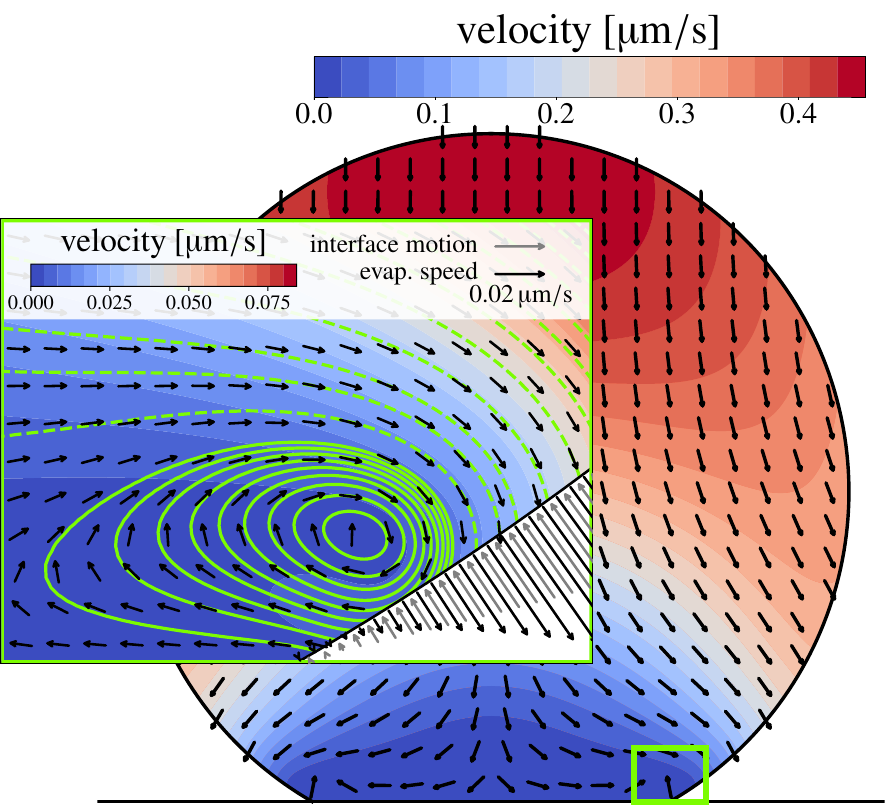}
  \caption{Isothermal numerical simulation of a \SI{100}{\nano\liter} water droplet with $\theta=\SI{150}{\degree}$ at \SI{20}{\celsius} and \SI{40}{\percent} relative humidity. Close to the pinned contact line, a flow inversion is visible, which grows for larger contact angles and vanishes for $\theta\lesssim\SI{133.4}{\degree}$. Color-coded is the velocity, stream lines are also shown in the inset (solid: positive, dashed: negative stream function iso-lines). The flow inversion results from the fact that the normal fluid velocity due to evaporation approaches zero faster than the motion of the interface when approaching the contact line. This is indicated by the arrows in the gas phase (black: evaporation velocity, grey: interface motion).}
  \label{fig:num}
\end{figure}

The velocity fields inside an evaporating droplet\cite{Masoud:2009} or liquid line\cite{Petsi:2006} of arbitrary contact angle have been derived under the simplifying assumption of inviscid flow and hence free slippage over the solid substrate. Stokes flow solutions have been obtained for either a uniform\cite{Petsi:2008} or a regularized\cite{Masoud:2009Stokes} evaporative flux. Similarity solutions to the Stokes flow subject to the evaporative flux (\ref{eq:flux2}) in the wedge-shaped region close to the contact line have been derived by Gelderblom et al.\,\cite{Gelderblom:2012}. This analysis showed that, surprisingly, for large contact angles ($\theta>127^\circ$) reversing flow structures appear. Moreover, for $\theta>133.4^\circ$ Moffatt eddies\cite{Moffatt:1964} dominate the flow near the contact line. To investigate how these corner flows relate to the flow in the entire droplet, numerical simulations are required. Up to now, however, numerical studies\cite{Fischer:2002,Hu:2005p522} focussed on smaller contact angles. In Fig.\,\ref{fig:num} we show, for the first time, numerical results for the velocity field inside a water droplet with a contact angle of 150$^\circ$. The inset indeed confirms the presence of the analytically predicted\cite{Gelderblom:2012} reversing corner flow. The extend of this region grows for higher contact angles and indeed vanishes at a critical contact angle of $\theta\approx 133.4^\circ$, however, the size of the region and the reversed velocity is small compared to the size and typical velocity of the entire droplet.

Experimentally, too, assessment of the flow inside a droplet on a hydrophobic substrate is challenging, as the shape of the droplet obscures the view in the contact-line area. Up to now, there is no experimental method available to resolve the predicted flows \cite{Petsi:2006,Petsi:2008,Masoud:2009,Masoud:2009Stokes, Gelderblom:2012} for evaporating droplets with contact angles above $90^\circ$.

 \section{Evaporation-driven Marangoni flow: the role of the liquid-gas interface}

Until now, the liquid-gas interface has been considered shear-free. However, in practice, the free surface is often  subject to an interfacial shear stress that gives rise to an additional internal flow. In this section, we will discuss how gradients in the interfacial temperature and in the concentration of surface-active material can give rise to interfacial shear, and, as a consequence, induce a Marangoni flow inside the evaporating droplet. 

Such a flow can be quantified by the Marangoni number $Ma$. In its most generic version, the $Ma$ number can be defined as the ratio between Marangoni-driven advective flow and the diffusive transport of the property generating the surface tension gradient (temperature or interfacial concentration of solute): $Ma = UL/D_X$, where $L$ is the length scale of the concentration gradient and $D_X$ is a diffusion constant of substance $X$ (with units of area per unit time). The order of magnitude of the Marangoni flow $U$ can be approximated by $\Delta \gamma/\mu$, where $\mu$ is the liquid's dynamic viscosity. The change of surface tension $\Delta \gamma$ can be expressed as a function of the surface property $X$, either the temperature or the concentration of solute along the interface, resulting in the following generic expression of the Marangoni number:
\begin{equation}\label{eq:Ma1}
    Ma = \left|\frac{d\gamma}{dX}\right|\frac{\Delta X L}{\mu D_X}.
\end{equation}
Most elements in the expression (\ref{eq:Ma1}) are liquid properties, and therefore known a priori, except for $\Delta X$ and $L$. In the following sections we will discuss some examples in which these quantities can be approximated. In particular, we will discuss important characteristics of Marangoni flows generated by thermal gradients (section \ref{sec:thermalmarangoni}) and by solutal gradients (section \ref{sec:solutalmarangoni}).


\subsection{Thermal Marangoni flow}\label{sec:thermalmarangoni}


Up to now, we described the evaporation process as \emph{thermodynamically} out-of-equilibrium, but \emph{thermally} in equilibrium. However, the loss of enthalpy in evaporating droplets unavoidably leads to a loss of thermal energy, manifested by a temperature decrease. Since the evaporative flux from the liquid-gas interface is non-uniform, it gives rise to an interfacial temperature gradient. As surface tension depends on temperature, this interfacial temperature gradient causes a gradient in surface tension and hence an interfacial shear. The flow generated inside the droplet by such  a thermal instability is known as \emph{thermal Marangoni flow}. 

The direction of the interfacial shear and hence the Marangoni flow is given by the direction of the temperature gradient along the liquid-gas interface \cite{Ristenpart:2007}. The strength of the Marangoni flow depends on several factors, and can be quantified by the Marangoni number (\ref{eq:Ma1}), which in the case of thermal gradients can be approximated a priori as: 
\begin{equation}\label{eq:Ma2}
    Ma = \left|\frac{d\gamma}{dT}\right|\frac{\rho H_vL}{\mu k},
\end{equation}
where $H_v$ is the latent heat of vaporization, $\mu$ is the liquid dynamic viscosity, and $k$ is the liquid's thermal conductivity. The variation in surface tension with temperature $d\gamma/dT$ is often termed the \emph{temperature coefficient}\cite{Navascues1979}, and takes negative values for pure liquids under atmospheric conditions for thermodynamic consistence. \cite{Navascues1979,deGennes1985wetting} 

\subsubsection{Direction of the Marangoni circulation}

As discussed above, the direction of the thermal Marangoni flow in evaporating droplets is determined by the direction of the interfacial temperature gradient. The energy required for the phase change is transferred to the interface through the liquid, the gas and the solid phases. Since the thermal conductivity of the solid phase is often orders of magnitude larger than that of the gas phase, solid-liquid heat transfer typically dominates the energy transport. The interfacial temperature gradient therefore strongly depends on the properties of the liquid droplet and the solid substrate\cite{Ristenpart:2007,Dunn:2009}.


For example, in the classical case of a sessile water droplet on a glass substrate, the solid phase is 1.6 times more conductive than the liquid phase. For droplets with a contact angle $<90^\circ$, the heat loss is largest at the contact line. Since glass is more thermally conductive than water, however, the heat is supplied from the solid substrate. Since the contact-line region is in direct contact with the substrate it will be the warmest region. As a result, a thermal gradient is established from the apex to the droplet base \cite{Ristenpart:2007}.
This effect was analyzed numerically by Diddens et al.~\cite{diddens2017evaporating}, who noticed that not only thermal conductivity of the solid and liquid is important, but also the substrate thickness, as suggested earlier by experiments and theoretical models \cite{Dunn:2008p62,Dunn:2009}. 

Interestingly, theoretical modelling predicts that interfacial temperature gradients can change directions (towards contact line or away from it) depending in a very subtle way on the liquid properties, droplet contact angle, substrate conductivity and evaporative cooling rate \cite{Ristenpart:2007,Nguyen:2018}. Such an inversion of the surface temperature gradient should result in an inversion of the Marangoni flow circulation. Unfortunately, up to now, no direct experimental evidence of such a flow inversions has been shown.

\subsubsection{Strength of the Marangoni flow}

The thermal Marangoni flow adds to the capillary flow discussed in \S 2, and it is directly proportional to the Marangoni number (\ref{eq:Ma2}).
Liquids with high volatility (large $H_v$) and low thermal conductivity (low $k$) yield large Marangoni numbers since they can support large thermal gradients along the interface. For these liquids, the internal flow is strongly dominated by thermal Marangoni flow. This strong Marangoni flow has for example been observed in ethanol (on highly conductive substrates), in early experiments with infrared imaging \cite{Brutin:2011hf}, and later confirmed in numerical simulations \cite{Karapetsas:2012ex} and experiments in micro-gravity conditions \cite{Carle:2012go}. 

 A case that deserves an extended discussion is that of water under atmospheric conditions. For a millimetric water droplet at room temperature (where $\left|d\gamma/dT\right|=$ 0.166 $\times10^{-3}$ \si{\newton\per \meter\per\kelvin}, $H_v=$2.26$\times10^6$ \si{\joule\per\kilogram} and $k=$0.59 \si{\joule\per\meter\per\second\per\kelvin})\cite{lide2004crc}, the Marangoni number reaches values of the order of $10^5$ \cite{Gelderblom:2012, Hu2005Marangoni}. 
Quantitative theoretical and numerical solutions for the Marangoni flow were first computed for sessile water droplets in atmospheric conditions in the limit of small contact angles by Hu \& Larson \cite{Hu2005Marangoni}. All numerical models following in the literature \cite{saenz2015evaporation, diddens2017evaporating,diddens2017modeling} predict that thermal Marangoni flow should overcome the capillary flow driving the coffee-stain effect by orders of magnitude. Only close to the contact line and on partially wetting substrates, where the evaporative flux diverges (see \S 2), the capillary flow dominates the Marangoni flow. Hence, a cross-over length scale should exist beyond which the Marangoni flow dominates \cite{Gelderblom:2012}. For most practical cases, this length-scale is sub-micron, which implies that the Marangoni flow should significantly alter the flow inside evaporating droplets.

Experimental observations, however, dramatically diverge from these theoretical and numerical findings. The disagreement between experimental and numerical values for the flow velocity reaches up to three orders of magnitude \cite{Xu:2007je,Trantum:2013ff,Marin2016surfactant,Rossi:2019}: from typical values of 1 \si{\micro\meter\per\second}
experimentally measured at room temperature conditions \cite{Marin2016surfactant,Rossi:2019}, up to 1 \si{\milli\meter\per\second}  in numerical simulations\cite{Hu2005Marangoni}. 
Similar conclusions were reached in early works on droplet evaporation by analyzing the heat transport within the evaporating droplet by laser interferometry \cite{BarnesHunter1982} and by temperature measurements \cite{Cammenga:1984}. In these studies it was found that thermal advection was negligible, despite the large Marangoni numbers computed.

\begin{figure}
    \centering
    \includegraphics[width=\columnwidth]{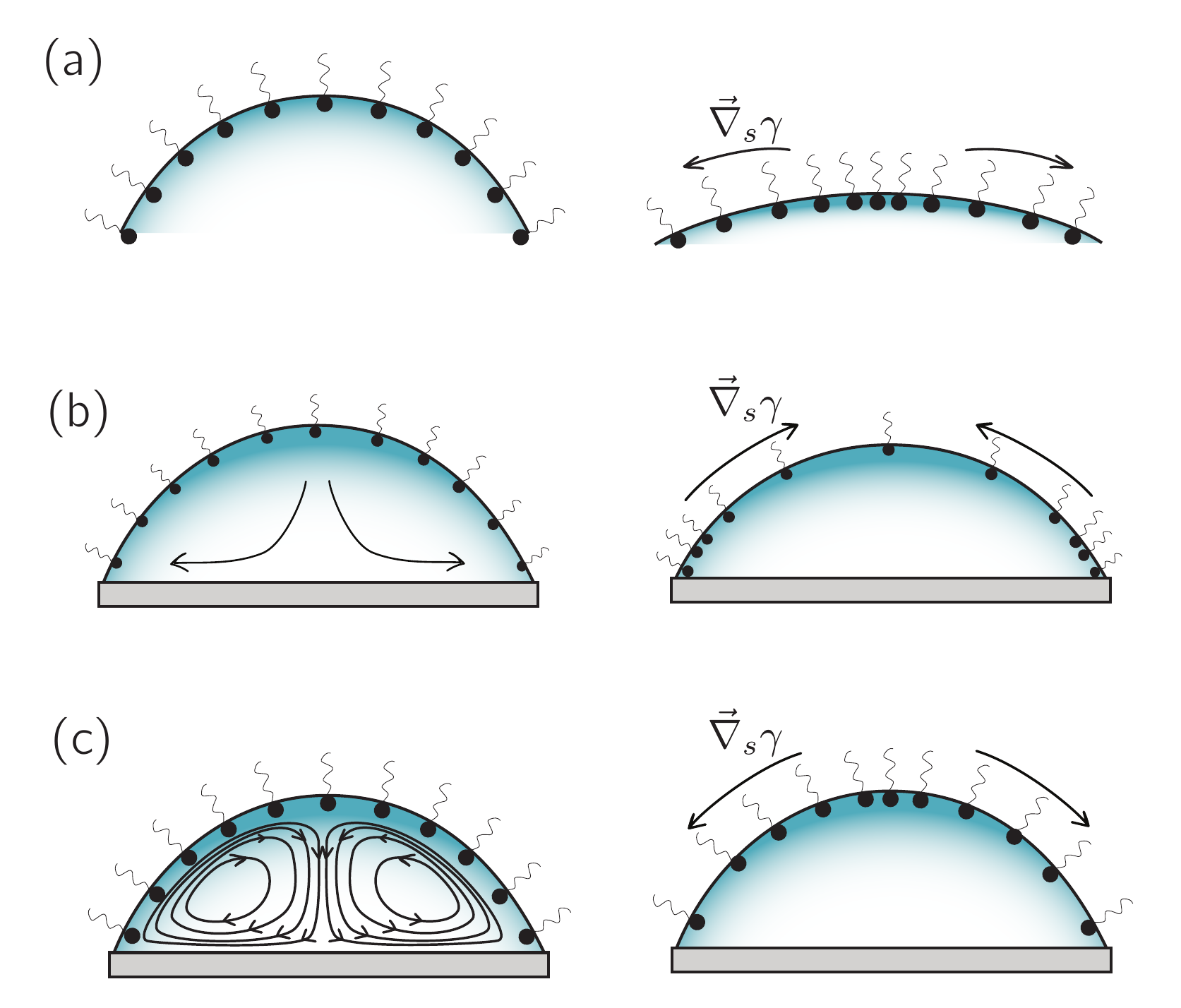}
    \caption{Three scenarios in which surfactants can generate an interfacial-concentration (and hence surface-tension) gradient in an evaporating droplet: (a) In a quiescent evaporating droplet, a concentration gradient is generated simply by the reduction of surface area. (b) In the presence of a capillary flow, surfactant accumulates in the vicinity of the contact line, generating a Marangoni stress directed towards the apex. (c) In the presence of a strong thermal interfacial Marangoni flow towards the apex, a solutal concentration gradient directed towards the contact line appears. Such solutal concentration gradient can partially neutralize the thermal Marangoni flow.}
    \label{fig:surfactants}
\end{figure}

Such disagreement between experiments and theory resembles the one found  in classical thermocapillary convection studies \cite{pearson1958convection}. In these systems, the discrepancy has been explained by the accidental presence of monolayers of surfactants (i.e.~contaminants). Surfactants can have a significant effect on the convective flow when they form a loose monolayer and can completely neutralize the flow when they form a compact monolayer (see the sketch in Fig. \ref{fig:surfactants})c, with surface concentrations of approximately $10^8$ and $10^{10}$ molecules per \si{\micro\meter\squared} respectively \cite{BergAcrivos1965}.
Cammenga et al. \cite{Cammenga:1984} and Hu \& Larson \cite{Hu2005Marangoni} also ascribed the discrepancy observed in evaporating droplets to certain amount of unknown contamination. 

Although the identity of these impurities is not yet known, their presence at the water-air interface has been clearly detected in the past. 
For example, making use of oscillating liquid bridges, Ponce-Torres, Vega \& Montanero \cite{Ponce2016Impurities} monitored the surface tension of different liquids as the liquid interface ``ages''. They obtained a minor decrease of surface tension for alcanes, and a substantial decrease for deionized water, from 40\% to 30\% depending on the atmosphere (open, saturated air or argon).
In a more recent example,  Molaei et al.\cite{molaei2021interfacial} studied the flow field generated by colloidal particles trapped at a quiescent water-air interface while experiencing Brownian motion. They obtained a surface-divergence free flow field, which is a clear signature of an interfacial incompressibility caused by contamination. The authors estimated an interfacial contaminant concentration of $10^3$ molecules/\SImum$^2$ and Marangoni numbers up to $Ma\approx 400$. Such estimates are compatible with the calculations by Hu \& Larson \cite{Hu2005Marangoni}. 

A recent numerical study confirms that surfactants or contaminants that reduce the surface tension of pure water by $\sim\SI{0.1}{\percent}$ are sufficient to suppress the thermal Marangoni flow \cite{van2022marangoni}. To our knowledge, no study has attempted yet to identify which contaminant(s) can be so generally present at water-air interfaces and induce practically the same disturbances in hundreds of different experiments for decades around the globe.

We cannot close this section without mentioning that water droplets far from atmospheric conditions can develop strong Marangoni flows, as evidenced in the extensive work of Ward et al. \cite{Ward:2004dg,Sefiane:2007he,Ghasemi:2010ge}. By carefully designing a closed setup to reduce the contamination of the liquid phase and manipulating the chamber's temperature and pressure, the authors reported interfacial velocities reaching up to 1 \si{\milli\meter\per\second}. Unfortunately, no comparison with numerical or analytical models was reported by the authors or by any posterior work.

\subsection{Solutal Marangoni flow}\label{sec:solutalmarangoni}

In the previous section we have discussed how surfactants can stabilize evaporation-driven flows in water droplets and other liquids. However, surfactants can also have the opposite effect and induce a surface flow when their dynamics couples with another destabilizing process such as evaporation or dissolution \cite{BergAcrivos1965,Nguyen:2002dda}. 
The simplest scenario that can be considered is an initially homogeneous distribution of insoluble surface active material at the liquid-gas interface of an evaporating droplet with a pinned contact line, as illustrated in Fig.~\ref{fig:surfactants}a. If the surface shrinks with a negligible surface velocity (tangential to the surface), surfactants will accumulate at the apex of the droplet due to the surface area reduction. As a consequence, a surface concentration gradient will be created that that points towards the apex, where the surface tension is then reduced. The resulting surface-tension gradient then drives a Marangoni flow towards the contact line. 

This simple (but unrealistic) case shows how easily the surfactant and the evaporation dynamics can be coupled to set a quiescent interface into motion. The situation becomes more complex in the case of soluble surfactants, which are found more often in practical situations. They can lead to a wide range of possible scenarios, from very dynamic and compressible interfaces (Figs. \ref{fig:surfactants}b and c) to surfactant-saturated and incompressible interfaces, depending on the cohesive/repulsive surfactant interaction, the time scales of the surfactant adsorption/desorption \cite{Stebe1996surfactantphyschem}, combined with the time scale of the evaporation process and the time scale of the surface flow \cite{Nguyen:2002dda}.

In the following section, we discuss a number of experimental studies on flow measurements performed in surfactant-laden droplets. However, since different physical processes are taking place in the droplet simultaneously and in competition (capillary flow, thermal Marangoni flow, solutal Marangoni flow, etc), fluid flow measurements are essentially incomplete and cannot give a complete picture on the role of surfactants \cite{Squires2020surfactant}. Therefore, such studies should combine experiments with numerical simulations to gain a deeper understanding of the dynamics of such complex systems.

\subsubsection{Experimental results}


Experimental measurements of the velocity field inside evaporating droplets with surfactants are scarce, but fortunately they do cover a variety of different surfactants. Using solutions of the popular soluble anionic surfactant sodium dodecyl sulfate (SDS) above the critical micelar concentration, Still et al.  \cite{still2012surfactant} observed the formation of a complex unsteady liquid flow, with an eddy in the vicinity of the contact line (a situation sketched in Fig. \ref{fig:surfactants}b). In a completely different setting, Sempels et al. \cite{Sempels:1bi} observed the formation of a similar kind of eddies in an evaporating sessile droplet containing \emph{P. aeruginosa} bacteria, which produce a substantial amount of bio-surfactants. The authors could reproduce similar flow structures using Triton X-100, another non-ionic surfactant. 

These results were confirmed and directly measured by Marin et al. \cite{Marin2016surfactant} using three-dimensional particle tracking. A characteristic Marangoni eddy was found systematically in SDS for concentrations ranging from the critical micellar concentration (CMC) up to 50$\times$CMC (as sketched in Fig. \ref{fig:surfactants}b). At extremely high concentrations, even an additional (but weaker) counter-rotating eddy could be identified and quantified.  Marin et al. \cite{Marin2016surfactant} also studied the effect of the non-ionic surfactant Polysorbate 80 (P-80), which is much larger and slower than SDS or Triton X-100. While SDS and Triton X-100 generated certain motion at the surface, P-80 causes the opposite effect: decreasing interfacial flow for increasing P-80 concentrations and an almost stagnant interface for bulk concentrations above the CMC.

\subsubsection{Numerical results}

The consideration of surfactants in numerical simulations of evaporating droplets requires a sophisticated treatment of the surfactant concentration field at the curved, moving and shrinking interface and its coupling to the bulk field via ad- and desorption processes. Recently, however, simulations of evaporating droplets containing insoluble \cite{karapetsas2016marangoni,van2020marangoni} and soluble surfactants \cite{van2021marangoni} have been accomplished.
In this latest work, the surface tension is assumed to follow a Frumkin equation of state with the surfactant concentration, taking into account steric interactions among the surfactants. The surfactant concentration tends to a dynamic equilibrium between the population at the interface and the population in the bulk. The formation of micelles in the bulk is also included by a dynamic equilibrium balance, which is activated when the bulk concentration reaches the CMC. The results of van Gaalen et al. \cite{van2021marangoni} agree qualitatively with experiments for bulk surfactant concentrations below the critical micellar concentration, confirming that the flow profiles found in the experiments \cite{Still:2012vj,Sempels:1bi,Marin2016surfactant} are caused by surfactants. For concentrations above the CMC, numerical simulations predict a weakening of the Marangoni-driven flow due to the increasing dominance of micelles in the system, which are assumed to be surface inactive. Interestingly, experimental results contradict such predictions: the experimental Marangoni-driven flow increases and becomes more and more complex as the bulk concentration increases above the CMC \cite{Sempels:1bi,Marin2016surfactant}. Such a disagreement is not surprising, given the complexity involved in evaporating sessile droplets, with both interfacial compression and shear, leading to regions of and high surfactant concentrations and others depleted of surfactant, combined with liquid convection in the bulk. 
Clearly, there is a the need for more sophisticated surfactant models that include surfactant interactions, heterogeneity, phase separation, etc \cite{Squires2020surfactant} to explain the experimental findings.


 \section{Evaporation-driven flows in mixtures}

In previous section we have discussed the role of amphiphilic substances, such as surfactants, in evaporation-driven flows in sessile droplets. However, also lyophobic substances (polar organic molecules, if the main solvent is water) and lyophilic substances (e.g. another solvent, or ionic salts) can have a crucial influence on the evaporation-driven flow. This influence can be twofold: in mixtures an interfacial (Marangoni) flow can arise due to gradients in solutal composition, and a buoyancy-driven natural convective flow can arise due to differences in density. Below we discuss both phenomena.
 
\subsection{Interfacial flow due to compositional gradients in mixtures}
In mixtures where the components have different vapor pressures, the preferential evaporation of one component induces a compositional gradient: on a partially wetting substrate the concentration of the most volatile component is lowest at the contact line (where the evaporation rate is largest). If the components also have a different surface tension, this preferential evaporation leads to a surface tension gradient, and hence, a Marangoni flow. 

The corresponding solutal Marangoni number can be estimated according to \eqref{eq:Ma1} by using a compositional quantity, e.g. the mass fraction $w$ of one component at the interface, for the surface property $X$. However, the estimation of the compositional difference $\Delta X=\Delta w$ in \eqref{eq:Ma1} involves the calculation of the evaporation rates of all constituents. Since these strongly depend on the local composition of the mixture due to Raoult's law, the vapor diffusivity of each component and the contact angle of the droplet, the interested reader is referred to Diddens et al. \cite{diddens2021competing} for a detailed derivation of the solutal Marangoni number for mixtures.

Perhaps the most well-known example of such a coupling between evaporation dynamics and solutal composition is the case of water and ethanol, featured in the \emph{tears of wine} phenomenon, in which the faster evaporation of ethanol at the triple line of a glass of wine drives a solutal Marangoni flow of liquid creeping up the glass \cite{thomson1855tears}. 


Evaporating sessile droplets of water and ethanol exhibit an erratic flow with larger fluctuations in the early stages, when the ethanol concentration is high, which relaxes to a classical capillary flow profile after certain amount of time, when the concentration of ethanol in the droplet is minimal \cite{christy2011flow,bennacer2014vortices}. This sequence of flow patterns have been reproduced using fully three-dimensional numerical simulations by Diddens et al. \cite{diddens2017evaporating}, since the Marangoni instability unavoidably breaks the axial symmetry of the system.

The same sequence of flow patterns are observed in popular alcoholic drinks such as ouzo \cite{Tan:2016kpa,diddens2017evaporating} or whisky \cite{KimWhisky2016}. Evaporating droplets of ouzo, a ternary system compound of water, ethanol and anise oil, also show such fluctuating flow in the first stage, until most of the ethanol is evaporated and the ternary system turns unstable, manifested in the nucleation of oil droplets \cite{Tan:2016kpa,diddens2017evaporating}.

\begin{figure}
    \centering
    \includegraphics[width=0.85\columnwidth]{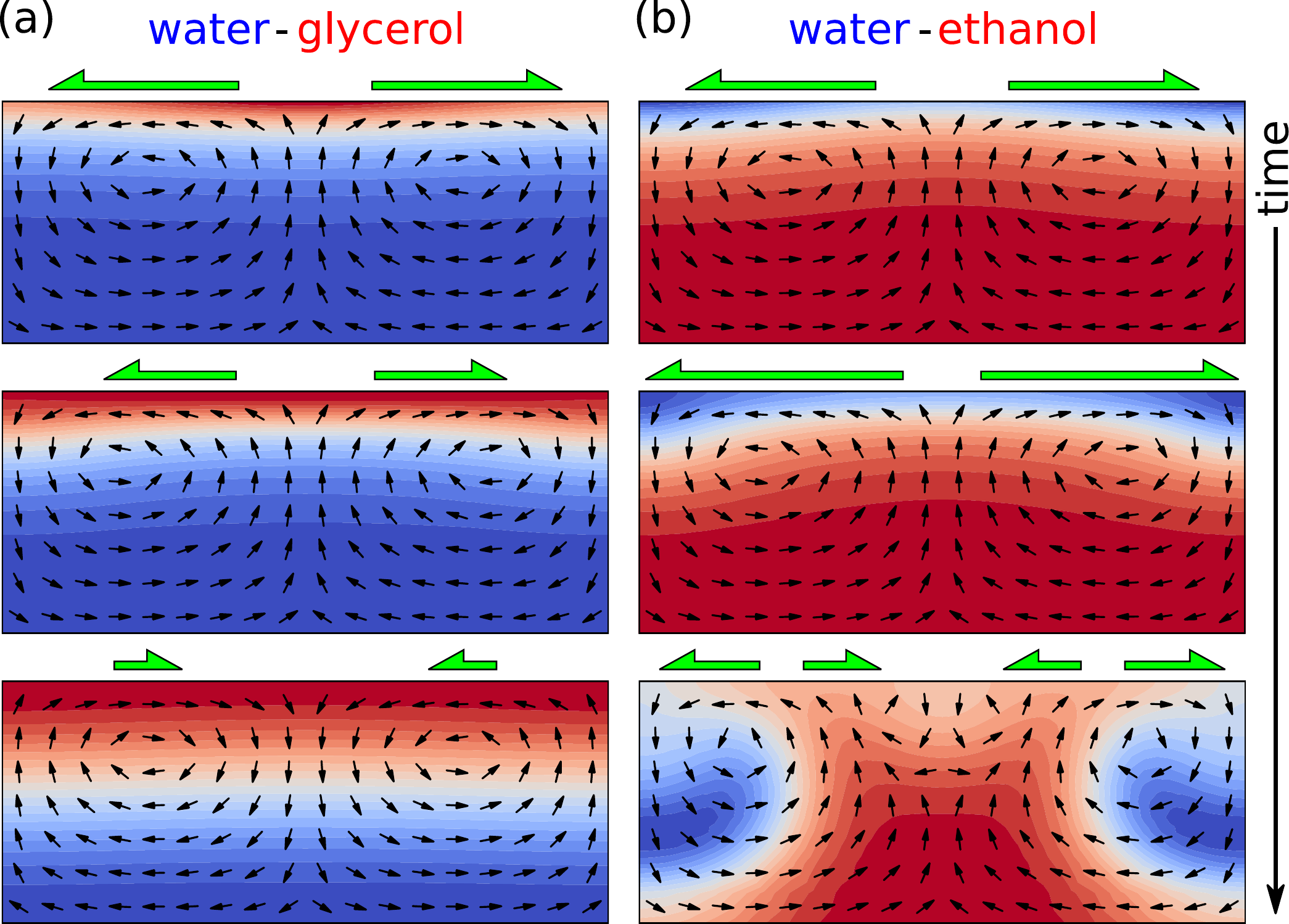}
    \caption{Different Marangoni dynamics depending on the volatility and the surface tension of the components. (a) In a water-glycerol system, water evaporates faster, i.e. enhanced glycerol concentrations (red) can be found near the the interface. A small compositional perturbation at the interface (top panel) is self-inhibiting, since the Marangoni flow (green arrows) pulls liquid with higher surface tension (water, blue) up from the bulk to the interface, to those areas with lowest surface tension (middle panel), resulting in a perturbation decay (bottom panel). (b) For an ethanol-water system, ethanol is more volatile. The smallest perturbation (top panel) leads to Marangoni flow that pulls up liquid with the lower surface tension (ethanol, red) and thereby enhancing the surface tension gradient (middle panel), leading to an instability and chaotic flow (bottom panel).}
    \label{fig:marainstab}
\end{figure}

The resulting flow when two or more components are left to evaporate in a sessile droplet can be quite different from the case of ethanol and water. A factor to observe is whether the most volatile liquid decreases the total
surface tension of the solution. If this is the case, as in the ethanol-water mixture, the resulting flow is typically largely dominated by a violent Marangoni surface flow that eventually breaks the axial symmetry of the system \cite{christy2011flow,bennacer2014vortices,diddens2017evaporating}. If the most volatile liquid increases the surface tension of the solution, as in a glycerol-water solution, the resulting Maragoni flow is a much weaker, gentle and axisymmetric flow. The reason for these entirely different flow scenarios can be found in the Marangoni instability,\cite{pearson1958convection,Sternling1959} which is explained in Fig.~\ref{fig:marainstab}. As we will see in the next section, in glycerol-water mixtures the density difference can have a more important influence on the internal flow than the stable Marangoni flow. 

Ethanol or glycerol are polar molecules which can be considered as lyophobic solutes that decrease the surface tension of water-based solutions. On the other hand, lyophilic solutes such as ionic salts or sucrose remain in the bulk and tend to increase surface tension, as is explained in \cite{hunter2001foundations}. Consequently, even at small concentrations they can have a tremendous impact on evaporation-driven flows. Marin et al. \cite{Marin2019Salt} showed that small amounts of sodium chloride can lead to a radical change of the flow structure inside the droplet. The evaporation-induced enrichment of such a solute at the droplet's contact line leads to a surface Marangoni flow towards the contact line. This Marangoni flow is the dominant source of motion within the droplet, over both the capillary and the thermal Marangoni flow \cite{Marin2019Salt}.

\subsection{Internal natural convection in mixtures} 

Typically, the size of the droplets considered in this review are comparable or below the capillary length. This means that their shape is dominated by surface tension and not by gravity (i.e.~the Bond number is much smaller than unity, see \S \ref{sec:partialwetting}).
However, gravitational effects can still have a significant influence on the fluid flow in an evaporating drop that consists of a mixture of two components with different densities. If the density difference is large enough, natural convection comes into play and contributes to the internal flow. 
The most paradigmatic example of natural convection in an evaporating droplet is the case of glycerol-water solutions \cite{Yaxing2019Glycerol}, due to the large density difference involved (glycerol is 25\% heavier than water). A similar flow has been observed in water/butanol and water/ethanol solutions by Edwards et al. \cite{Edwards2018Gravity}, although in the latter case solutal Marangoni flows enter in competition with natural convection. 


In principle, both Marangoni instabilities and natural convection might be present in any arbitrary binary solution. Assuming that thermal effects can be neglected and that the base capillary flow is much weaker than such instabilities, Diddens et al. \cite{diddens2021competing} could reduce the problem using two dimensionless numbers, namely the Marangoni number (eq. \ref{eq:Ma1}) and the Rayleigh number (the ratio between transport via natural convection and diffusion). For a more elaborated discussion on the different regimes that can be found, we refer the interested reader to Diddens et al.\cite{diddens2021competing}.


\section{Particle transport and deposition by evaporation-driven flows}\label{sec:particles}
When an evaporating droplet contains tiny particles, these will get transported by the internal flow and deposit onto the substrate, leaving a stain. The patterns formed by these drying stains are a signature of the evaporation dynamics\cite{Sefiane:2010}. In this section we will describe step by step how the capillary and Marangoni flows discussed in Sections 2 and 3 affect the particle deposition. We will restrict ourselves to the hydrodynamic transport of neutrally buoyant colloidal particles only, and to dilute suspensions, in which particle-particle interaction and changes in the suspension rheology during evaporation are negligible. 
There are very few situations in which only hydrodynamics can determine the final fate of the dispersed particles. In general, the final position of a dispersed particle in an evaporating droplet involves complex physico-chemical processes that will not be discussed in depth in this review. 

In evaporating drops two modes of particle transport exist: transport via Brownian motion of the particles and transport by convection in the evaporation-driven flow\cite{Petsi:2010}. The ratio of importance of these two modes is expressed by the Peclet number
\begin{equation}
    \mathrm{Pe}=\frac{UR_p^2}{L D_p}\label{eq:peclet}
\end{equation}
with $U$ the characteristic velocity scale, $R_p$ the particle radius, $L$ is a characteristic length scale for convective transport (e.g.~the droplet radius or initial inter-particle spacing) and $D_p=k_BT/6\pi \mu R_p$ the particle diffusion constant as expressed via the Stokes-Einstein relation. Note that, as discussed in section \ref{sec:pinned}, the flow field within the droplet is rather heterogeneous, and therefore the Peclet number can take different values in different regions within the same droplet. 

When $\mathrm{Pe}\ll 1$ the particle diffusive timescale is much shorter than the convective timescale, and particles will cross streamlines instead of following them. By contrast, when $\mathrm{Pe}\gg 1$ particles will get transported by the flow and follow the streamlines because of their hydrodynamic drag\cite{Petsi:2010}. In that case, the internal flow will have a crucial impact on the particle deposition. 

We will discuss the influence the internal flow on particle deposition patterns for droplets with a pinned contact line where capillary flow dominates in \S \ref{sec:pincap} and for a dominant Marangoni flow in \S \ref{sec:pinmar}. The deposition pattern for droplets where the contact line moves continuously or shows stick-slip behavior is described in \S \ref{sec:clmotion}.

\subsection{Pinned contact lines and capillary flow}\label{sec:pincap}

In a droplet evaporating with a pinned contact line, the capillary flow causes the deposition of particles into a ring-shaped stain\cite{Deegan:1997, deegan2000pattern}: the well-known \emph{coffee-stain effect}. This particle deposition by capillary flows is so robust and reproducible that it has generated an immense interest and led to many applications\cite{Han:2012}. 

\subsubsection{Ring-shaped stain formation}
\begin{figure}[ht]
\centering
  \includegraphics[width=.8\columnwidth]{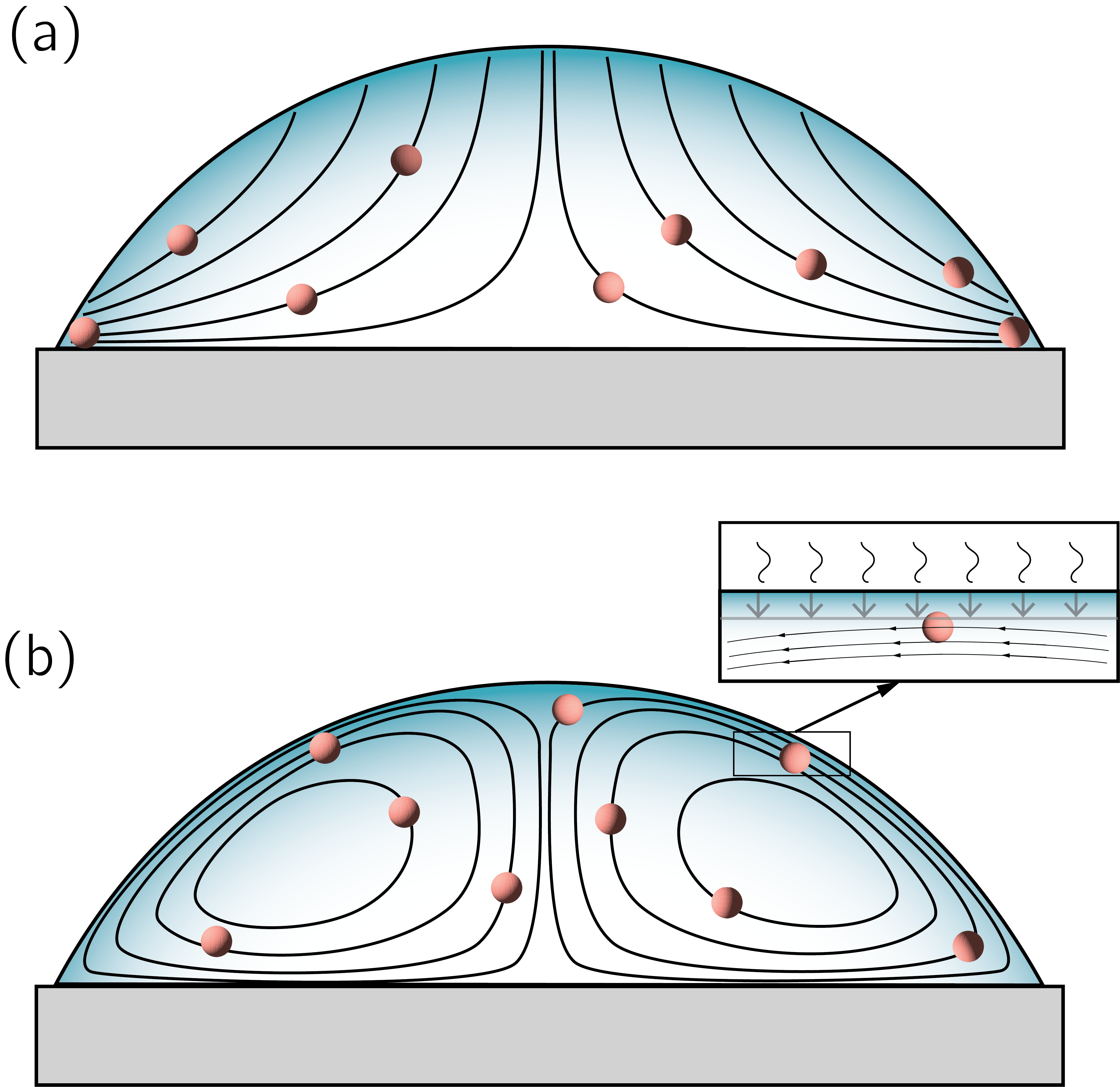}
  \caption{(a) The `canonical' capillary flow is characterized by open streamlines that start at the liquid-gas interface and end up at the contact line. Consequently, every particle in the volume will always end up at the contact line. (b) When other flow sources appear in the system, the streamlines typically close and -hydrodynamically speaking- the particles do not have a predestined end location. The final location of a particle will depend on the physico-chemical forces that come into play when it approaches boundaries, specially of the receding liquid-gas interface (see insert).}
  \label{fig:streamlines}
\end{figure}

Clearly, the ring-shaped stain is a consequence of the fact that, for pinned droplets with a small contact angle ($\theta<127^\circ$)\cite{Gelderblom:2012} {in thermal equilibrium, {the velocity field in every point within the droplet points radially outwards towards the contact line}\cite{Deegan:1997}, and consequently all streamlines that start at the liquid-air interface end there (Fig. \ref{fig:streamlines}a).} 
Hence, all particles following the streamlines in principle end up at the contact line. As discussed above, particles follow the streamlines of the flow when their Peclet number is much larger than unity. For a capillary flow, the Peclet number (\ref{eq:peclet}) is given by
\begin{equation}
    \mathrm{Pe}=\frac{\Delta c}{\rho} \frac{D}{D_p}\frac{R_p^2}{RL\theta_0}\frac{1}{\left(1-t/t_f\right)\sqrt{1-r/R}}\label{eq:peclet2}
\end{equation}
where we used (\ref{eq:avuapp}) for the characteristic capillary velocity with $\theta(t)$ given by (\ref{eq:theta}), $\theta_0$ is the initial contact angle, and $t_f$ is the total lifetime of the droplet. Hence, for small contact angles and/or close to the contact line of the droplet $\mathrm{Pe}\gg 1$ and particle convection dominates diffusion. However, even for $\mathrm{Pe}\gg 1$  there are three effects that can prevent particles from ending up at the contact line. 

First of all, particles could get adsorbed at the solid substrate before they reach the contact line, due to a combination of hydrodynamic and physico-chemical effects: Particles will not remain on a single stationary streamline for the entire droplet lifetime but follow the instantaneous streamlines that change in time as the droplet shrinks. When a particle ends up on a streamline passing close to the solid substrate, it can eventually get adsorbed either by simple mechanical friction, electrostatic \cite{NogueraMarin:2015} or chemical (e.g.~Van der Waals) forces\cite{Butt1991}. Hence, non-hydrodynamic forces, i.e.~the physico-chemistry of the solvent, particle surface and boundary material, eventually determine whether the particle can continue its path or will be adsorbed at the boundary. Clearly, adsorption becomes even more important if the Peclet number
(eq.~\ref{eq:peclet2}) is much smaller than unity and the particle motion is dominated by Brownian motion. In that case particles can spontaneously change streamline, and eventually attach irreversibly to the solid substrate before reaching the contact line.   
 
 Second, particles could get adsorbed at the receding liquid-gas interface.
Any particle moving on a streamline parallel to a receding liquid-gas interface (for example when there are closed streamlines, as in Fig.\,\ref{fig:streamlines}b), will eventually come close to the interface. Note that this is again a purely hydrodynamic (and geometric) fact, in which the physico-chemistry has not yet been invoked. 
Similarly to the particle-solid interaction described above, as soon as the radius of the particle becomes smaller than the distance to the interface, non-hydrodynamic forces come into play, i.e.~Van der Waals forces, electrostatics and wetting \cite{Kaz:2011df}. In principle, such hydrodynamic particle trapping could also occur for open streamlines \cite{Jafari2016} (see Fig.\,\ref{fig:streamlines}a), i.e.~in thermal equilibrium and for a force-free interface. 
Hence, regardless of the streamline configuration, hydrodynamics tells us that particles dispersed in an evaporating drop will approach the liquid-gas interface at a rate proportional to its receding motion. However, their final destination ultimately depends on the physico-chemistry of both the particle and the boundary.


Third, the particles may not have enough time to reach the contact line before the droplet has completely dried \cite{Kaplan:2015}. To estimate the ratio between the particle transport time and the droplet drying time we will use the results of \S \ref{sec:pinned} for droplets with a small initial contact angle $\theta_0\ll 90^\circ$. The total drying time of the droplet scales as (see equation (\ref{eq:theta})) $t_{f}\sim\theta_0 \rho R^2/ D\Delta c$, with $\theta_0$ the initial contact angle of the droplet. For the convective timescale we take the time required for a particle to move from a position $r$ to the contact line $t_c\sim (R-r)/U$ with $U$ the characteristic height-averaged velocity (\ref{eq:avuapp}). Hence
\begin{equation}
    \frac{t_c}{t_f}=\frac{\theta(t)}{\theta_0}\left(1-r/R\right)^{3/2}=\left(1-\frac{t}{t_f}\right)\left(1-r/R\right)^{3/2}.\label{eq:dryingtime}
\end{equation}
From (\ref{eq:dryingtime}) one observes that at early times $t\ll t_f$ and for particles far from the contact line ($r\ll R$) the time needed to reach the contact line is comparable to the total drying time of the droplet ($t_c/t_f=1$). By contrast, particles close to the contact line ($r\approx R$) always quickly end up in the ring stain. 
Note that the situation where $t_c/t_f>1$ (droplet dries before particles have reached the contact line) is impossible, due to the fact that the internal flow is directly driven by the evaporation: if the evaporation is faster, so is the flow transporting the particles. At late times ($t\sim t_f$), the ratio $t_c/t_f\approx 0$ and convection is fast compared to the total drying time of the droplet. Hence, eventually most particles will end up in the ring stain, except for that fraction adsorbed at the solid substrate (as discussed above, mostly for $\mathrm{Pe}\ll 1$). 

Experimentally, the ring-shaped stain pattern found in droplets with pinned contact lines is remarkably robust and reproducible \cite{Conway:1997vd,Deegan:1997,Kajiya:2008hz,Berteloot:2012dn,Boulogne2016drywet}, in the sense that other factors as particle size or substrate material are subdominant to the capillary flow transporting the particles. Importantly, the accumulation of particles at the contact line due to the capillary flow reinforces the pinning of the contact line (a phenomenon termed self-pinning\cite{deegan2000pattern,Weon2013selfpinning}), thereby increasing the robustness of the ring stain. With the droplet's contact line pinned and therefore static, contact-line dynamics has no influence on the deposition pattern, which results in a reproducible ring stain.

\subsubsection{Width of the ring stain}
The rate at which the ring stain grows depends on time, due to the time-dependence of the velocity field inside the droplet (\ref{eq:avu}). To calculate the mass and width of the ring stain over time, we follow the argument by Deegan\cite{deegan2000contact,deegan2000pattern}. First, we calculate the time it takes for the capillary flow to transport all particles located between a radial position $r=R-d$, where $d\ll R$, and $r=R$ towards the contact line (i.e.~towards $r=R$, see Fig.~\ref{fig:class}) and find\cite{deegan2000contact}
\begin{equation}
    t=\int_{R-d}^R \frac{\mathrm{d}r}{\overline{u}(r)}\sim \frac{\theta \rho \sqrt{R}}{D\Delta c}d^{3/2},\label{eq:width1}
\end{equation}
where we used the approximate expression (\ref{eq:avuapp}) for the height-averaged velocity. From (\ref{eq:width1}) we obtain that $d\sim t^{2/3}$. As long as $d\ll R$ (i.e.~the early-time regime), the volume swept towards the contact line in time $t$ scales as\cite{deegan2000pattern} $\theta R d^2=\theta^{-1/3} R^3 t^{4/3} (D\Delta c/\rho R^2)^{4/3}$. The resulting mass of the ring stain that forms is then given by 
\begin{equation}
    m_R(t)\sim \Phi_0 \theta Rd^2\sim\Phi_0 \theta^{-1/3} R^3 t^{4/3}\left(\frac{D\Delta c}{\rho R^2}\right)^{4/3}, \label{eq:width2}
\end{equation}
with $\Phi_0$ the initial particle concentration. The corresponding volume of the stain is given by $V_s\sim m_R(t)/\rho_p R_p^3 \Phi_f=\theta R w^2$, with $\Phi_f$ the packing fraction of the stain, $\rho_p$ the particle density and $w$ the stain width. Hence, the width of the stain (i.e.~the property that one can easily observe during and after an experiment) scales as
\begin{equation}
    \frac{w(t)}{R}\sim \frac{1}{\sqrt{\rho_pR_p^3}}\left[\frac{D\Delta c}{\rho \theta R^2}\right]^{2/3}\left(\frac{\Phi_0}{\Phi_f}\right)^{1/2}t^{2/3}.\label{eq:stainwidth}
\end{equation}
From (\ref{eq:stainwidth}) one observes that the width of the stain initially grows as a power law in time\cite{deegan2000contact}. At late times a time-divergent behaviour is observed\cite{deegan2000contact,Marin:2011} and (\ref{eq:stainwidth}) no longer applies as the criterion $d\ll R$ is violated. The increase in growth rate is further reinforced by the fact that the packing fraction of the stain is not a constant. Indeed, as the evaporation proceeds, the packing fraction of the stain decreases\cite{Marin:2011,marin2012building}. This decrease in packing fraction is a direct consequence of the temporal divergence in the particle velocity: at early times, when the deposition speed of particles arriving at the contact line is low, they have time to arrange themselves by Brownian motion into an ordered crystal (i.e.~with a high packing fraction), as illustrated in Fig.~\ref{fig:introfig}c. By contrast, at late times particles arrive at high speed and are jammed into a disordered phase\cite{Marin:2011} (i.e.~low packing fraction). This balance between the timescale for Brownian motion and the (rapidly decreasing) convective timescale can be expressed by a time-dependent Peclet number (\ref{eq:peclet2}). Here, the relevant convective length scale is given by the typical spacing between the particles,  which depends on the particle concentration. For dilute suspensions, $L\sim \Phi_0^{-1/3}$\cite{Marin:2011}. 
  When $\mathrm{Pe}(t)\ll 1$ (hence early times $t\ll t_f$) a crystalline structure forms, whereas $\mathrm{Pe}(t)\gg 1$ (late times $t\sim t_f$) particles get jammed into a disordered phase\cite{Marin:2011}. 

The scaling analysis above is restricted to the case where $\theta\ll 90^\circ$, due to the simplified expressions used for e.g.~the internal velocity field. On hydrophobic substrates, however, similar effects will occur in the later stages of the droplet life: when the contact line remains pinned, the droplet will always pass through a regime where $\theta\ll 90^\circ$ and the majority of the ring stain forms.  


\subsection{Pinned contact lines and Marangoni flow}\label{sec:pinmar}

Whenever the capillary flow encounters a competing source of flow, most of the streamlines --that previously ended at the static contact line-- will now be closed, typically forming eddies distributed radially along the droplet volume (see Fig. \ref{fig:streamlines}b). These eddies can occur both due to the presence of forces (Marangoni stresses) at the liquid-gas interface or by volumetric forces (natural convection), as discussed in previous sections. Particles following closed streamlines will be simply circulating within the droplet volume until \emph{something} stops them. As the droplet volume shrinks, the chances that a particle gets intercepted by the liquid-gas interface, or gets attached to the substrate increase. Nonetheless, from a purely hydrodynamic perspective, one cannot immediately tell where the particles will end up. In the lines below, we will discuss the conditions under which one can make a valid prediction on the particles' fate. 

The simplest situation would occur if the particles are lyophilic (i.e.~have no tendency to breach the liquid-gas interface) and have low affinity for the solid substrate. In that case, the particles will be simply circulating all over the volume until the solvent is gone. The expected result should be an homogeneous distribution of particles along the droplet's contact area. Note that the nature of the competing flow is not important: thermal\cite{Hu:2006}, solutal \cite{majumder2012overcoming} Marangoni flows or natural convective flows should give similar results. Clearly, in the absence of a hydrodynamic deposition mechanism, non-hydrodynamic forces become more important to determine the final position of the particles. For example, the addition of surface-adsorbed macromolecules to enhance the particle adhesion to the substrate has been shown to promote homogeneous deposition of particles in an evaporating droplet with a dominant Marangoni flow.  \cite{KimWhisky2016}

Another situation in which the particles destination can be predicted, even when the streamlines are closed, occurs when particles are adsorbed at the liquid-air interface. The fate of the particles depends then on the direction of the interfacial flow, and the final pattern depends on the particle concentration. For example, in droplets where a solutal Marangoni flow is directed from the apex to the contact line, particles can get adsorbed at the liquid-air interface and transported towards the contact line, forming a ring-shaped stain. In contrast to the classical 3D ring-shaped stains formed by the capillary flow \cite{Deegan:1997,Marin:2011}, this ring would actually be formed at the liquid-gas interface and is therefore two-dimensional \cite{Marin2019Salt,Bruning:2020ej}. 
When particles adsorb at the interface and the interfacial flow is directed towards the droplet apex, particles would instead accumulate around the apex, either forming a ``Marangoni ring'' or a cap, depending on the strength of the flow. This is typically observed with thermal Marangoni flows \cite{Deegan:1997,Rossi:2019,Parsa2015,Lijun2022}, but the final fate of the particles agglomerated at the droplet apex  depends on the particle affinity to the liquid interface, the interfacial/bulk flow and the receding interface.
Nevertheless, when particle concentration is high enough and/or particles have a strong affinity for the interface \cite{bigioni2006kinetically,Yunker:2011vr}, particles form a dense network that immobilises the interface and leading to an homogeneous deposition of particles.
The presence of interfacial Marangoni flows increases the time particles spend in the vicinity of the interface and therefore increases the chances of adsorption. Nevertheless, note that hydrodynamics play a secondary role in this scenario, and the colloidal physico-chemistry is dominant. While there is some degree of understanding on the dominating forces attaching colloids to solid surfaces \cite{Harting2018friction}, interfacial adsorption is a complex issue under debate and strongly dependent on the particle surface chemistry and on the solvent's properties \cite{Kaz:2011df,Ballard2019ColloidsInterfaces,Isa2021particlesinterfaces}. Consequently, homogeneous distributions of particles cannot be achieved by exploiting hydrodynamics only.

We would like to stress that in none of the described situations in this subsection the capillary flow completely disappears. Depending on the strength of the dominant recirculating flow, a region of variable size will always remain in the vicinity of the contact line in which the capillary flow still dominates\cite{Gelderblom:2012}. This capillary flow will transport a certain amount of particles transported towards the contact line, which is manifested by a thin line that denotes the contact line position \cite{Hu:2006,Ristenpart:2007, Still:2012vj, malinowski2018dynamic}.

\subsection{Moving contact lines and stick-slip behavior}\label{sec:clmotion}


 In pure liquids, the contact-line dynamics is governed by capillary forces (i.e.~the unbalanced Young's force) and the interaction with the substrate.\cite{Orejon:2011} In suspensions however, the presence of particles induces a self-pinning behaviour:\cite{deegan2000pattern,Weon2013selfpinning} the particles deposited at the contact line due to the capillary flow provide an additional force\cite{Zhang:2014} that keeps the contact line pinned. When the contact angle becomes small enough, the contact line is pulled away from the deposit by the unbalanced Young's force and recedes.\cite{deegan2000pattern} Often, local depinning occurs and the contact line switches between pinned and depinned states (termed stick-slip\cite{Rio:2006} or stick-slide\cite{Stauber:2015} behaviour), leaving behind a trace of heterogeneous deposits.\cite{deegan2000pattern,Rio:2006} The strength of these self-pinning events depends on the time the contact line can remain pinned and deposit can accumulate. The pinning time depends on the substrate wettability, manifested in large contact angles (i.e.~more space in the wedge-shaped geometry that needs to be filled with particles in order to pin) and larger roughness or contact-angle hysteresis\cite{NogueraMarin:2014} (i.e.~a large difference between advancing and receding angle, which makes it easier to pin the contact line, and hence increases the pinning time).

In the presence of stick-slip motion of the contact line, when the radially outward capillary flow dominates the flow, a pattern of concentric rings is typically left behind. Such a pattern depends on properties such as the evaporation rate, particle concentration and intrinsic viscosity. \cite{Frastia:2012} However, if the capillary flow is sub-dominant, the self-pinning effect is typically weaker. Additionally, if the particles do not have a strong affinity for the substrate, the final result is a smooth contact line motion (no stick-slip visible). In that case, most particles will remain in suspension until the droplet evaporates completely, leaving a compact stain at the center of the droplet, much smaller than the initial droplet contact diameter. \cite{mampallil2012control}

 In evaporating droplets with large contact angles, i.e.~on hydrophobic and superhydrophobic substrates, the contact line typically recedes smoothly. In these cases, the patterns left by evaporating suspension droplets depend on the contact-line dynamics and the amount of particles in the droplet, rather than on the hydrodynamics of the internal flow. Such systems can yield either flat particle agglomerates, when the particle number is small \cite{seyfert2021evaporation}, or three-dimensional spheroids, when the particle number is large \cite{Rastogi:2008ba,Marin:2012bw,seyfert2021evaporation}.

\section{Open Questions and Future Directions}


In this review we presented our current understanding on evaporation-driven flows in sessile droplets. As we have seen, these tiny and seemingly mundane systems present complicated and surprising physical challenges and have, and will continue having, a great impact on many fields. We hope that our work will help a wide community of scientists to discover (or to relearn) the  beautiful science contained in tiny vanishing liquid volumes. Twenty-five years after the appearance of its foundational paper\cite{Deegan:1997}, many open questions in the field still remain. Moreover, new challenges appeared over the years, for example through the discovery of active colloids. To conclude our review, we now summarize the most prominent outstanding questions.
\begin{itemize}
\item
\emph{Evaporation-driven flows outside the partial wetting regime:} Very soon after one leaves the classical configuration of a sessile droplet with pinned contact line and small contact angle, internal flows become hard to determine, both in experiment and in theory.
On the one hand, on wetting surfaces with vanishing contact angles a coupling arises between the evaporative flux, internal flow, droplet shape and contact-line dynamics that complicates theoretical progress. On the other hand, on hydrophobic surfaces the droplet geometry poses challenges to both experimental measurements as well as theoretical calculations of the internal flow. 
Previous asymptotic analyses and the numerical results of the present paper have revealed intriguing reversing flow structures arise at large contact angle that certainly deserve more attention, but would require the development of new experimental techniques to overcome the optical hurdles involved. The impact of these flow structures (if any) on particle deposition patterns also remains to be explored. 

\item \emph{Contact-line singularities:} In the contact-line region of an evaporating droplet extremely complicated phenomena are at play, and one quickly encounters the limitations of a continuum description. Even for the classical situation of a pinned droplet on a partial wetting substrate, the physical mechanisms that regularize contact-line singularities in both the evaporative flux and the internal flow field are still far from understood. Progress in this direction is unavoidably linked with the debate on contact-line dynamics \cite{Snoeijer:2013}. Accurate modelling of what happens at the contact line requires a coupling between the macroscopic and the -yet to be clearly defined- microscopic phenomena happening at the contact line in an evaporating droplet.

\item \emph{Surfactants:}
As we have seen, interfacial-driven flows in water-based droplets are typically two or three orders of magnitude smaller than those obtained by numerical simulations or theoretical predictions. Moreover, while this issue has been first identified in the context of thermal Marangoni flows,  it also extends to solutal Marangoni flows. The most accepted explanation for these observations is the presence of contamination at water-air interfaces. However, the physical chemistry and origin of such ubiquitous contaminants still remain unknown and require further experimental investigation. The confirmation of such a contamination effect might lead to a better understanding of surfactant dynamics. The development of accurate surfactant-dynamics models is also essential for further numerical progress on (solutal) Marangoni flows. 

\item \emph{Particle physical-chemistry \& deposition patterns:}
The majority of the literature one can find nowadays on evaporation-driven flows in droplets is motivated by the dream of controlling the patterns that emerge from the non-volatile material within the droplets. This control is often claimed to be achieved by manipulating the hydrodynamic flow. In this context, the only hydrodynamic mechanism capable of achieving a reproducible and general (independent on the type of particle and  solvent employed) control on particle deposition is the pure radially outward capillary flow described by Deegan et al.\cite{Deegan:1997} As discussed in \S\ref{sec:particles}, this is the only evaporation-driven flow in which all streamlines end at the contact line, dragging particles towards the contact line in an universal way, independent of their nature. The presence of any additional source of flow unavoidably leads to a closure of the streamlines, and, consequently, the fate of the dispersed particles cannot be determined solely by hydrodynamics. Instead, the particle deposition pattern becomes strongly dependent on the particle physical chemistry, i.e.~their adsorption to the solid substrate and to the liquid-air interface, 
and will be highly specific for the particular system (particles, solvent, substrate) under study. 
 
Probably the most effective way of obtaining an homogeneous distribution of particles with pinned contact lines is to have particles adsorb strongly to the liquid-air interface as it recedes, such that an interfacial monolayer of particles is formed. \cite{bigioni2006kinetically,Yunker:2011vr} A few recipes are known to obtain such strong interfacial adhesion, but a proper understanding of the phenomenon and broader exploration of means to achieve it is still lacking. 


\item {\emph{Evaporation-driven rheological changes:}
In \S \ref{sec:particles} we have discussed particle deposition in evaporation-driven flow. In our discussion, we always considered the case where the suspension remains dilute up to the moment that the particles get deposited and leave the suspension. In doing so, we have not discussed the complications arising when the non-volatile material that remains in suspension reaches high concentrations during droplet evaporation. 

For large particle concentrations, the suspension rheology gets altered. These changes in the suspension's rheological response can be highly localized e.g.~at the contact line, where the evaporation is strongest and the particle transport largest \cite{Rio:2006}. The altered rheology can affect both the internal flow and the motion of the contact line and thereby give rise to peculiar deposition patterns\cite{Thiele:2014}, such as regularly spaced concentric rings\cite{Frastia:2012, Eren:2021}, striped, fingered or branched structures \cite{Berteloot2012,Weon2013PRE}.  

Moreover, a dense contact-line deposit can alter the local flow, which now encounters a porous medium\cite{Mahadevan2015} with a porosity dependent on the particle size, which in turn affects the deposition pattern. Such models do not consider solutes of different nature, and a modification to take these differences into account would be desirable. How close such models can reproduce the experimentally observed patterns for solid particles, and whether their packing can be predicted, is yet a problem to tackle. A detailed exploration of how both rheological changes and the porous structure of large particles inside a deposit modify the internal flow, contact line motion and thereby give rise to a plethora of different deposition patterns is still lacking. Moreover, a detailed description of the dynamic particle ordering inside the deposit requires a coupling between the macroscopic dense-suspension flow and the microscopic deposition dynamics\cite{Mahadevan2015}.}

\item \emph{Self-propelled colloids:}
The discovery of active colloidal particles has opened a myriad of possibilities  \cite{IntroActiveMatter2020,practicalguide2020}, not only from the engineering point of view, but for the study of natural systems \cite{poon2013clarkia}. The presence of synthetic or biological self-propelled particles inside an evaporating-driven flow involves two competing time scales, the particle's own self-propelling time-scale and the droplet's flow time-scale, which lead to a rich and novel scenario. On top of that, hydrodynamic interactions between the active swimmers could give rise to an active stress that locally alters the flow inside an evaporating droplet\cite{Kasyap:2014}. However, the unavoidable presence of surfactants, salts or other solvent mixtures in such solutions induce solutal Marangoni flows that interfere with the flows under study. Therefore, to understand the interaction between the convective flow and the particle's own propulsion the droplet's evaporation-driven flow needs to be well characterized \cite{Sempels:1bi,Callegari2019bacteriaring}. 

A specially interesting subclass of systems yet to be explored is that of active particles at interfaces \cite{Paolo2016activeinterface,dietrich2017activeinterface} of evaporating droplets. If the interfacial flow is weak enough, such particles would explore the droplet interface at their own time scale, independent of the bulk flow.
\end{itemize}





\section{Acknowledgements}
The authors are grateful to Stephen Wilson, Jacco Snoeijer and Detlef Lohse for stimulating discussions. HG acknowledges financial support from the Netherlands Organisation for Scientific Research (NWO) through Veni Grant No.~680-47-451.
AM acknowledges financial support from the European Research Council Starting Grant No. 678573.



\begin{thebibliography}{100}

\bibitem{Callegari2019bacteriaring}
T.~Andac, P.~Weigmann, S.~K.~P. Velu, E.~Pinçe, G.~Volpe, G.~Volpe, and
  A.~Callegari.
\newblock Active matter alters the growth dynamics of coffee rings.
\newblock {\em Soft Matter}, 15:1488--1496, 2019.

\bibitem{Ballard2019ColloidsInterfaces}
N.~Ballard, A.~D. Law, and S.~A.~F. Bon.
\newblock Colloidal particles at fluid interfaces: behaviour of isolated
  particles.
\newblock {\em Soft Matter}, 15:1186--1199, 2019.

\bibitem{BarnesHunter1982}
G.~Barnes and D.~Hunter.
\newblock Heat conduction during the measurement of the evaporation resistances
  of monolayers.
\newblock {\em J. Colloid Interface Sci.}, 88(2):437--443, 1982.

\bibitem{bennacer2014vortices}
R.~Bennacer and K.~Sefiane.
\newblock Vortices, dissipation and flow transition in volatile binary drops.
\newblock {\em J. Fluid Mech}, 749(5):649--665, 2014.

\bibitem{BergAcrivos1965}
J.~Berg and A.~Acrivos.
\newblock The effect of surface active agents on convection cells induced by
  surface tension.
\newblock {\em Chem. Eng. Sci.}, 20(8):737--745, 1965.

\bibitem{Berteloot:2012dn}
G.~Berteloot, A.~Hoang, A.~Daerr, H.~P. Kavehpour, F.~Lequeux, and L.~Limat.
\newblock {Evaporation of a sessile droplet: Inside the coffee stain}.
\newblock {\em J. Colloid Interface Sci.}, 370(1):155--161, Mar. 2012.

\bibitem{Berteloot2012}
G.~Berteloot, A.~Hoang, A.~Daerr, H.~P. Kavehpour, F.~Lequeux, and L.~Limat.
\newblock Evaporation of a sessile droplet: Inside the coffee stain.
\newblock {\em J. Colloid Interface Sci.}, 370(1):155--161, 2012.

\bibitem{Berteloot:2008}
G.~Berteloot, C.~T. Pham, A.~Daerr, F.~Lequeux, and L.~Limat.
\newblock Evaporation-induced flow near a contact line: consequences on coating
  and contact angle.
\newblock {\em Europhys. Lett.}, 83:14003, 2008.

\bibitem{bigioni2006kinetically}
T.~P. Bigioni, X.-M. Lin, T.~T. Nguyen, E.~I. Corwin, T.~A. Witten, and H.~M.
  Jaeger.
\newblock Kinetically driven self assembly of highly ordered nanoparticle
  monolayers.
\newblock {\em Nat. Mater.}, 5(4):265--270, 2006.

\bibitem{Bocquet:2007}
L.~Bocquet.
\newblock Tasting edge effects.
\newblock {\em Am. J. Phys.}, 75:148, 2007.

\bibitem{bodiguel2013imaging}
H.~Bodiguel and J.~Leng.
\newblock Imaging the drying of a colloidal suspension: Velocity field.
\newblock {\em Chem. Eng. Process.: Process Intensification}, 68:60--63, 2013.

\bibitem{Bonn:2009}
D.~Bonn, J.~Eggers, J.~Indekeu, J.~Meunier, and E.~Rolley.
\newblock Wetting and spreading.
\newblock {\em Rev. Mod. Phys.}, 81(2):739--805, 2009.

\bibitem{Boulogne:2013}
F.~Boulogne, F.~Giorgiutti-Dauphin{\'e}, and L.~Pauchard.
\newblock The buckling and invagination process during consolidation of
  colloidal droplets.
\newblock {\em Soft Matter}, 9(3):750--757, 2013.

\bibitem{Boulogne2016drywet}
F.~Boulogne, F.~Ingremeau, and H.~A. Stone.
\newblock Coffee-stain growth dynamics on dry and wet surfaces.
\newblock {\em J. Physics: Condens. Matter}, 29(7):074001, dec 2016.

\bibitem{Bourouiba:2021}
L.~Bourouiba.
\newblock The fluid dynamics of disease transmission.
\newblock {\em Annu. Rev. Fluid Mech.}, 53:473--508, 2021.

\bibitem{Bruning:2020ej}
M.~A. Bruning, L.~Loeffen, and A.~Marin.
\newblock {Particle monolayer assembly in evaporating salty colloidal
  droplets}.
\newblock {\em {Phys. Rev. Fluids}}, pages 1--18, Aug. 2020.

\bibitem{Brutin:2011hf}
D.~Brutin, B.~Sobac, F.~Rigollet, and C.~Le~Niliot.
\newblock {Infrared visualization of thermal motion inside a sessile drop
  deposited onto a heated surface}.
\newblock {\em Exp. Therm. Fluid Sci.}, 35(3):521--530, 2011.

\bibitem{Butt1991}
H.-J. Butt.
\newblock Measuring electrostatic, van der waals, and hydration forces in
  electrolyte solutions with an atomic force microscope.
\newblock {\em Biophys. J.}, 60(6):1438--1444, 1991.

\bibitem{Cammenga:1984}
H.~Cammenga, D.~Schreiber, G.~Barnes, and D.~Hunter.
\newblock {On Marangoni convection during the evaporation of water}.
\newblock {\em J. Colloid Interface Sci.}, 98:585--586, 1984.

\bibitem{Carle:2012go}
F.~Carle, B.~Sobac, and D.~Brutin.
\newblock {Hydrothermal waves on ethanol droplets evaporating under terrestrial
  and reduced gravity levels}.
\newblock {\em J. Fluid Mech.}, 712:614--623, Nov. 2012.

\bibitem{Carrier:2016}
O.~Carrier, N.~Shahidzadeh-Bonn, R.~Zargar, M.~Aytouna, M.~Habibi, J.~Eggers,
  and D.~Bonn.
\newblock Evaporation of water: evaporation rate and collective effects.
\newblock {\em J.Fluid Mech.}, 798:774--786, 2016.

\bibitem{Cazabat:2010}
A.~M. Cazabat and G.~Gu{\'e}na.
\newblock Evaporation of macroscopic sessile droplets.
\newblock {\em Soft Matter}, 6(12):2591--2612, 2010.

\bibitem{Stebe1996surfactantphyschem}
J.~Chen and K.~J. Stebe.
\newblock {Marangoni Retardation of the Terminal Velocity of a Settling
  Droplet: The Role of Surfactant Physico-Chemistry}.
\newblock {\em J. Colloid Interface Sci.}, 178(1):144--155, Mar. 1996.

\bibitem{christy2011flow}
J.~R. Christy, Y.~Hamamoto, and K.~Sefiane.
\newblock Flow transition within an evaporating binary mixture sessile drop.
\newblock {\em Phys. Rev. Lett.}, 106(20):205701, 2011.

\bibitem{Conway:1997vd}
J.~Conway, H.~Korns, and M.~R. Fisch.
\newblock {Evaporation Kinematics of Polystyrene Bead Suspensions}.
\newblock {\em Langmuir}, pages 1--6, Feb. 1997.

\bibitem{IntroActiveMatter2020}
M.~Das, C.~F. Schmidt, and M.~Murrell.
\newblock Introduction to active matter.
\newblock {\em Soft Matter}, 16:7185--7190, 2020.

\bibitem{deGennes1985wetting}
P.-G. De~Gennes.
\newblock Wetting: statics and dynamics.
\newblock {\em Rev. Mod. Phys.}, 57(3):827, 1985.

\bibitem{Eren:2021}
A.~Dede~Eren, D.~Eren, T.~Wilting, J.~de~Boer, H.~Gelderblom, and J.~Foolen.
\newblock Self-agglomerated collagen patterns govern cell behaviour.
\newblock {\em Sci. Rep.}, 11:1516, 2021.

\bibitem{deegan2000pattern}
R.~D. Deegan.
\newblock Pattern formation in drying drops.
\newblock {\em Phys. Rev. E}, 61(1):475, 2000.

\bibitem{Deegan:1997}
R.~D. Deegan, O.~Bakajin, T.~F. Dupont, G.~Huber, S.~R. Nagel, and T.~A.
  Witten.
\newblock Capillary flow as the cause of ring stains from dried liquid drops.
\newblock {\em Nature}, 389(6653):827--828, 1997.

\bibitem{deegan2000contact}
R.~D. Deegan, O.~Bakajin, T.~F. Dupont, G.~Huber, S.~R. Nagel, and T.~A.
  Witten.
\newblock Contact line deposits in an evaporating drop.
\newblock {\em Phys. Rev. E}, 62(1):756, 2000.

\bibitem{diddens2017modeling}
C.~Diddens, J.~G.~M. Kuerten, C.~W.~M. Van~der Geld, and H.~M.~A. Wijshoff.
\newblock Modeling the evaporation of sessile multi-component droplets.
\newblock {\em J. Colloid Interface Sci.}, 487:426--436, 2017.

\bibitem{diddens2021competing}
C.~Diddens, Y.~Li, and D.~Lohse.
\newblock Competing marangoni and rayleigh convection in evaporating binary
  droplets.
\newblock {\em J. Fluid Mech.}, 914:A23, 2021.

\bibitem{diddens2017evaporating}
C.~Diddens, H.~Tan, P.~Lv, M.~Versluis, J.~G.~M. Kuerten, X.~Zhang, and
  D.~Lohse.
\newblock Evaporating pure, binary and ternary droplets: thermal effects and
  axial symmetry breaking.
\newblock {\em J. Fluid Mech.}, 823:470--497, 2017.

\bibitem{dietrich2017activeinterface}
K.~Dietrich, D.~Renggli, M.~Zanini, G.~Volpe, I.~Buttinoni, and L.~Isa.
\newblock Two-dimensional nature of the active brownian motion of catalytic
  microswimmers at solid and liquid interfaces.
\newblock {\em N. J. Phys.}, 19(6):065008, 2017.

\bibitem{Dunn:2008p62}
G.~J. Dunn, S.~K. Wilson, B.~R. Duffy, S.~David, and K.~Sefiane.
\newblock A mathematical model for the evaporation of a thin sessile liquid
  droplet: Comparison between experiment and theory.
\newblock {\em Colloids Surf. A}, 323(1-3):50--55, 2008.

\bibitem{Dunn:2009}
G.~J. Dunn, S.~K. Wilson, B.~R. Duffy, S.~David, and K.~Sefiane.
\newblock The strong influence of substrate conductivity on droplet
  evaporation.
\newblock {\em J. Fluid Mech.}, 623:329--351, 2009.

\bibitem{Edwards2018Gravity}
A.~M.~J. Edwards, P.~S. Atkinson, C.~S. Cheung, H.~Liang, D.~J. Fairhurst, and
  F.~F. Ouali.
\newblock Density-driven flows in evaporating binary liquid droplets.
\newblock {\em Phys. Rev. Lett.}, 121:184501, Nov 2018.

\bibitem{Eggers:2010}
J.~Eggers and L.~M. Pismen.
\newblock Nonlocal description of evaporating drops.
\newblock {\em Phys. Fluids}, 22(11):112101, 2010.

\bibitem{Erbil:2012}
H.~Erbil.
\newblock Evaporation of pure liquid sessile and spherical suspended drops: A
  review.
\newblock {\em Adv. Colloid Interface Sci.}, 170:67--86, 2012.

\bibitem{Fischer:2002}
B.~Fischer.
\newblock Particle convection in an evaporating colloidal droplet.
\newblock {\em Langmuir}, 18:60--67, 2002.

\bibitem{Frastia:2012}
L.~Frastia, A.~Archer, and U.~Thiele.
\newblock Modelling the formatipon of structures deposits at receding contact
  lines of evaporating solutions and suspensions.
\newblock {\em Soft Matter}, 8:11363, 2012.

\bibitem{Gelderblom:2012}
H.~Gelderblom, O.~Bloemen, and J.~H. Snoeijer.
\newblock Stokes flow near the contact line of an evaporating drop.
\newblock {\em J. Fluid Mech.}, 709:69--84, 2012.

\bibitem{Gelderblom:2011}
H.~Gelderblom, A.~G. Mar\'in, H.~Nair, A.~van Housselt, L.~Lefferts, J.~H.
  Snoeijer, and D.~Lohse.
\newblock How water droplets evaporate on a superhydrophic substrate.
\newblock {\em Phys. Rev. E}, 83(2):026306, 2011.

\bibitem{Ghabache:2016}
E.~Ghabache, G.~Liger-Belair, A.~Antkowiak, and T.~S\'{e}on.
\newblock Evaporation of droplets in a champagne wine aerosol.
\newblock {\em Sci. Reports}, 6(25148):25148, 2016.

\bibitem{Ghasemi:2010ge}
H.~Ghasemi and C.~Ward.
\newblock Energy transport by thermocapillary convection during
  sessile-water-droplet evaporation.
\newblock {\em Phys. Rev. Lett.}, 105(13):136102, 2010.

\bibitem{Guena:2007}
G.~Gu{\'e}na, C.~Poulard, and A.~M. Cazabat.
\newblock {The leading edge of evaporating droplets}.
\newblock {\em J. Colloid Interface Sci.}, 312(1):164--171, 2007.

\bibitem{Han:2012}
W.~Han and Z.~Lin.
\newblock Learning from ``coffee rings": ordered structures enabled by
  controlled evaporative self-assembly.
\newblock {\em Angew. Chem. Int. Ed.}, 51:1534--1546, 2012.

\bibitem{Hu2005Marangoni}
H.~Hu and R.~Larson.
\newblock {Analysis of the Effects of Marangoni Stresses on the Microflow in an
  Evaporating Sessile Droplet}.
\newblock {\em Langmuir}, 21(9):3972--3980, Apr. 2005.

\bibitem{Hu:2006}
H.~Hu and R.~Larson.
\newblock Marangoni effect reverses coffee-ring depositions.
\newblock {\em J. Phys. Chem. B}, 110(14):7090--7094, 2006.

\bibitem{Hu:2005p522}
H.~Hu and R.~G. Larson.
\newblock Analysis of the microfluid flow in an evaporating sessile droplet.
\newblock {\em Langmuir}, 21(9):3963--3971, 2005.

\bibitem{Huh:1971}
C.~Huh and L.~E. Scriven.
\newblock Hydrodynamic model of steady movement of a solid/liquid/fluid contact
  line.
\newblock {\em J. Colloid Interface Sci.}, 35(1):85--101, 1971.

\bibitem{hunter2001foundations}
R.~J. Hunter.
\newblock {\em Foundations of colloid science}.
\newblock Oxford university press, 2001.

\bibitem{Jafari2016}
S.~Jafari~Kang, V.~Vandadi, J.~D. Felske, and H.~Masoud.
\newblock Alternative mechanism for coffee-ring deposition based on active role
  of free surface.
\newblock {\em Phys. Rev. E}, 94:063104, Dec 2016.

\bibitem{Kajiya:2008hz}
T.~Kajiya, D.~Kaneko, and M.~Doi.
\newblock {Dynamical Visualization of Coffee Stain Phenomenon in Droplets of
  Polymer Solution via Fluorescent Microscopy}.
\newblock {\em Langmuir}, 24(21):12369--12374, 2008.

\bibitem{Kaplan:2015}
C.~N. Kaplan and L.~Mahadevan.
\newblock {Evaporation-driven ring and film deposition from colloidal
  droplets}.
\newblock {\em J. Fluid Mech.}, 781:R2--13, Sept. 2015.

\bibitem{Mahadevan2015}
C.~N. Kaplan and L.~Mahadevan.
\newblock Evaporation-driven ring and film deposition from colloidal droplets.
\newblock {\em J. Fluid Mech.}, 781:R2, 2015.

\bibitem{Karapetsas:2012ex}
G.~Karapetsas, O.~K. Matar, P.~Valluri, and K.~Sefiane.
\newblock {Convective Rolls and Hydrothermal Waves in Evaporating Sessile
  Drops}.
\newblock {\em Langmuir}, 28(31):11433--11439, Aug. 2012.

\bibitem{karapetsas2016marangoni}
G.~Karapetsas, K.~C. Sahu, and O.~K. Matar.
\newblock Evaporation of sessile droplets laden with particles and insoluble
  surfactants.
\newblock {\em Langmuir}, 32(27):6871--6881, 2016.

\bibitem{Kasyap:2014}
T.~Kasyap, D.~Koch, and M.~Wu.
\newblock Bacterial collective motion near the contact line of an evaporating
  sessile drop.
\newblock {\em Phys. Fluids}, 26:111703, 2014.

\bibitem{Kaz:2011df}
D.~M. Kaz, R.~McGorty, M.~Mani, M.~P. Brenner, and V.~N. Manoharan.
\newblock {Physical ageing of the contact line on colloidal particles at liquid
  interfaces.}
\newblock {\em Nat. Mater.}, 11(2):138--142, Feb. 2012.

\bibitem{KimWhisky2016}
H.~Kim, F.~Boulogne, E.~Um, I.~Jacobi, E.~Button, and H.~A. Stone.
\newblock Controlled uniform coating from the interplay of marangoni flows and
  surface-adsorbed macromolecules.
\newblock {\em Phys. Rev. Lett.}, 116(12):124501, 2016.

\bibitem{Kovalchuk:2014}
N.~Kovalchuk, A.~Trybala, and V.~Starov.
\newblock Evaporation of sessile droplets.
\newblock {\em Curr. Opin. Colloid Interface Sci.}, 19:336--342, 2014.

\bibitem{langmuir:1918}
I.~Langmuir.
\newblock The evaporation of small spheres.
\newblock {\em Phys. Rev.}, 12(368), 1918.

\bibitem{Larson:2014}
R.~Larson.
\newblock Transport and deposition patterns in drying sessile droplets.
\newblock {\em AlChE J.}, 60(5):1538--1571, 2014.

\bibitem{Lebedev}
N.~Ledebev.
\newblock {\em Special functions and their applications}.
\newblock Prentice-Hall, 1965.

\bibitem{Yaxing2019Glycerol}
Y.~Li, C.~Diddens, P.~Lv, H.~Wijshoff, M.~Versluis, and D.~Lohse.
\newblock Gravitational effect in evaporating binary microdroplets.
\newblock {\em Phys. Rev. Lett.}, 122:114501, Mar 2019.

\bibitem{lide2004crc}
D.~R. Lide.
\newblock {\em CRC handbook of chemistry and physics}, volume~85.
\newblock CRC press, 2004.

\bibitem{Loussert:2016}
C.~Loussert, A.~Bouchaudy, and J.-B. Salmon.
\newblock Drying dynamics of a charged colloidal dispersion in a confined drop.
\newblock {\em Phys. Rev. Fluids}, 1(8):084201, 2016.

\bibitem{majumder2012overcoming}
M.~Majumder, C.~S. Rendall, J.~A. Eukel, J.~Y. Wang, N.~Behabtu, C.~L. Pint,
  T.-Y. Liu, A.~W. Orbaek, F.~Mirri, J.~Nam, et~al.
\newblock Overcoming the ''coffee-stain'' effect by compositional
  marangoni-flow-assisted drop-drying.
\newblock {\em J. Phys. Chem. B}, 116(22):6536--6542, 2012.

\bibitem{Paolo2016activeinterface}
P.~Malgaretti, M.~N. Popescu, and S.~Dietrich.
\newblock Active colloids at fluid interfaces.
\newblock {\em Soft Matter}, 12:4007--4023, 2016.

\bibitem{malinowski2018dynamic}
R.~Malinowski, G.~Volpe, I.~P. Parkin, and G.~Volpe.
\newblock Dynamic control of particle deposition in evaporating droplets by an
  external point source of vapor.
\newblock {\em J. Phys. Chem. Lett.}, 9(3):659--664, 2018.

\bibitem{Mampallil:2018}
D.~Mampallil and H.~Eral.
\newblock A review on suppression and utilization of the coffee-ring effect.
\newblock {\em Adv. Colloid Interface Sci.}, 252:38--54, 2018.

\bibitem{mampallil2012control}
D.~Mampallil, H.~Eral, D.~Van Den~Ende, and F.~Mugele.
\newblock Control of evaporating complex fluids through electrowetting.
\newblock {\em Soft Matter}, 8(41):10614--10617, 2012.

\bibitem{Squires2020surfactant}
H.~Manikantan and T.~M. Squires.
\newblock Surfactant dynamics: hidden variables controlling fluid flows.
\newblock {\em J. Fluid Mech.}, 892:P1, 2020.

\bibitem{Marin:2011}
A.~Marin, H.~Gelderblom, D.~Lohse, and J.~H. Snoeijer.
\newblock Order-to-disorder transition in ring-shaped colloidal stains.
\newblock {\em Phys. Rev. Lett.}, 107:085502, 2011.

\bibitem{Marin2011Rush}
A.~Marin, H.~Gelderblom, J.~Snoeijer, and D.~Lohse.
\newblock Rush hour in evaporating coffee drops.
\newblock {\em Phys. Fluids}, 23(091111):1--2, 2011.

\bibitem{Marin:2012bw}
A.~Marin, H.~Gelderblom, A.~Susarrey-Arce, A.~van Houselt, H.~Gardeniers,
  D.~Lohse, and J.~H. Snoeijer.
\newblock {Building microscopic soccer balls with evaporating colloidal fakir
  drops}.
\newblock {\em Proc. Natl. Acad. Sci.}, 109(41):16455--16458, 2012.

\bibitem{Marin2019Salt}
A.~Marin, S.~Karpitschka, D.~Noguera-Mar{\'\i}n, M.~A. Cabrerizo-V{\'\i}lchez,
  M.~Rossi, C.~J. K{\"a}hler, and M.~A.~R. Valverde.
\newblock Solutal marangoni flow as the cause of ring stains from drying salty
  colloidal drops.
\newblock {\em Phys. Rev. Fluids}, 4(4):041601, 2019.

\bibitem{Marin2016surfactant}
A.~Marin, R.~Liepelt, M.~Rossi, and C.~J. K\"ahler.
\newblock Surfactant-driven flow transitions in evaporating droplets.
\newblock {\em Soft Matter}, 12(5):1593--1600, 2016.

\bibitem{marin2012building}
{\'A}.~G. Mar{\'\i}n, H.~Gelderblom, A.~Susarrey-Arce, A.~van Houselt,
  L.~Lefferts, J.~G. Gardeniers, D.~Lohse, and J.~H. Snoeijer.
\newblock Building microscopic soccer balls with evaporating colloidal fakir
  drops.
\newblock {\em Proc. Natl. Acad. Sci.}, 109(41):16455--16458, 2012.

\bibitem{Masoud:2009Stokes}
H.~Masoud and J.~Felske.
\newblock Analytical solution for stokes flow inside an evaporating sessile
  drop: Spherical and cylindrical cap shapes.
\newblock {\em Phys. Fluids}, 21:042102, 2009.

\bibitem{Masoud:2009}
H.~Masoud and J.~D. Felske.
\newblock Analytical solution for inviscid flow inside an evaporating sessile
  drop.
\newblock {\em Phys. Rev. E}, 79(016301):042102--1--042102--1, 2009.

\bibitem{maxwell:1877}
J.~Maxwell.
\newblock Diffusion, collected scientific papers.
\newblock In {\em Encyclopedia Britannica}. Cambridge, 1877.

\bibitem{Moffatt:1964}
H.~K. Moffatt.
\newblock Viscous and resistive eddies near a sharp corner.
\newblock {\em J. Fluid Mech.}, 18:1--18, 1964.

\bibitem{molaei2021interfacial}
M.~Molaei, N.~G. Chisholm, J.~Deng, J.~C. Crocker, and K.~J. Stebe.
\newblock Interfacial flow around brownian colloids.
\newblock {\em Phys. Rev. Lett.}, 126(22):228003, 2021.

\bibitem{Morris:2014}
S.~Morris.
\newblock On the contact region of a diffusion-limited evaporating drop: a
  local analysis.
\newblock {\em J. Fluid Mech.}, 793:308--337, 2014.

\bibitem{Navascues1979}
G.~Navascues.
\newblock Liquid surfaces: theory of surface tension.
\newblock {\em Rep. Prog. Phys.}, 42(7):1131--1186, jul 1979.

\bibitem{Nguyen:2018}
T.~Nguyen, S.~Biggs, and A.~Nguyen.
\newblock Analytical model for diffusive evaporation of sessile droplets
  coupled with interfacial cooling effect.
\newblock {\em Langmuir}, 34:6955--6962, 2018.

\bibitem{Nguyen:2002dda}
V.~Nguyen and K.~Stebe.
\newblock {Patterning of Small Particles by a Surfactant-Enhanced
  Marangoni-B{\'e}nard Instability}.
\newblock {\em {Phys. Rev. Lett.}}, 88(16):164501, Apr. 2002.

\bibitem{NogueraMarin:2014}
D.~Noguera-Mar{\'\i}n, C.~L. Moraila-Mart{\'\i}nez, M.~A.
  Cabrerizo-V{\'\i}lchez, and M.~A. Rodr{\'\i}guez-Valverde.
\newblock {Transition from Stripe-like Patterns to a Particulate Film Using
  Driven Evaporating Menisci}.
\newblock {\em Langmuir}, 30(25):7609--7614, July 2014.

\bibitem{NogueraMarin:2015}
D.~Noguera-Mar{\'\i}n, C.~L. Moraila-Mart{\'\i}nez, M.~A.
  Cabrerizo-V{\'\i}lchez, and M.~A. Rodr{\'\i}guez-Valverde.
\newblock {In-plane particle counting at contact lines of evaporating colloidal
  drops: effect of the particle electric charge}.
\newblock {\em Soft Matter}, 11:987--993, Jan. 2015.

\bibitem{Orejon:2011}
D.~Orejon, K.~Sefiane, and M.~Shanahan.
\newblock Stick-slip of evaporating droplets: Substrate hydrophobicity and
  nanoparticle concentration.
\newblock {\em Langmuir}, 27:12834--12843, 2011.

\bibitem{Oron:1997}
A.~Oron, S.~Davis, and S.~Bankoff.
\newblock Long-scale evolution of thin liquid films.
\newblock {\em Rev. Mod. Phys.}, 69:931--980, 1997.

\bibitem{Parsa2015}
M.~Parsa, S.~Harmand, K.~Sefiane, M.~Bigerelle, and R.~Deltombe.
\newblock {Effect of substrate temperature on pattern formation of
  nanoparticles from volatile drops}.
\newblock {\em Langmuir}, 31(11):3354--3367, 2015.

\bibitem{Pauchard:2004}
L.~Pauchard and Y.~Couder.
\newblock Invagination during the collapse of an inhomogeneous spherical shell.
\newblock {\em Europhys. Lett.}, 66(5):667, 2004.

\bibitem{pearson1958convection}
J.~Pearson.
\newblock On convection cells induced by surface tension.
\newblock {\em J. Fluid Mech.}, 4(5):489--500, 1958.

\bibitem{Petsi:2005}
A.~Petsi and V.~Burganos.
\newblock Potential flow inside an evaporating cylindrical line.
\newblock {\em Phys. Rev. E}, 72(4):047301, 2005.

\bibitem{Petsi:2006}
A.~Petsi and V.~Burganos.
\newblock Evaporation-induced flow in an inviscid liquid line at any contact
  angle.
\newblock {\em Phys. Rev. E}, 73(4):041201, 2006.

\bibitem{Petsi:2010}
A.~Petsi, A.~Kalarakis, and V.~Burganos.
\newblock Deposition of brownian particles during evaporation of
  two-dimensional sessile droplets.
\newblock {\em Chem. Eng. Sci.}, 65:2978--2989, 2010.

\bibitem{Petsi:2008}
A.~J. Petsi and V.~N. Burganos.
\newblock Stokes flow inside an evaporating liquid line for any contact angle.
\newblock {\em Phys. Rev. E}, 78:036324--1--036324--9, 2008.

\bibitem{Pham:2010}
C.~T. Pham, G.~Berteloot, F.~Lequeux, and L.~Limat.
\newblock Dynamics of complete wetting liquid under evaporation.
\newblock {\em Europhys. Lett.}, 92(5):54005, 2010.

\bibitem{Picknett:1977p108}
R.~G. Picknett and R.~Bexon.
\newblock The evaporation of sessile or pendant drops in still air.
\newblock {\em J. Colloid Interface Sci.}, 61(2):336--350, 1977.

\bibitem{Ponce2016Impurities}
A.~Ponce-Torres, E.~Vega, and J.~Montanero.
\newblock Effects of surface-active impurities on the liquid bridge dynamics.
\newblock {\em Exp. Fluids}, 57(5):1--12, 2016.

\bibitem{poon2013clarkia}
W.~Poon.
\newblock From clarkia to escherichia and janus: The physics of natural and
  synthetic active colloids.
\newblock {\em Proc. Int. Sch. Phys. Enrico Fermi}, 184:317--386, 2013.

\bibitem{Popov:2005}
Y.~O. Popov.
\newblock Evaporative deposition patterns: Spatial dimensions of the deposit.
\newblock {\em Phys. Rev. E}, 71(3):036313, 2005.

\bibitem{Poulard:2003}
C.~Poulard, O.~Benichou, and A.~M. Cazabat.
\newblock Freely receding evaporating droplets.
\newblock {\em Langmuir}, 19(21):8828--8834, 2003.

\bibitem{Poulard:2005b}
C.~Poulard, G.~Guena, and A.~Cazabat.
\newblock Diffusion-driven evaporation of sessile drops.
\newblock {\em J. Phys.: Condens. Matter}, 17:S4213--S4227, 2005.

\bibitem{Poulard:2005}
C.~Poulard, G.~Gu{\'e}na, A.~M. Cazabat, A.~Boudaoud, and M.~{Ben Amar}.
\newblock Rescaling the dynamics of evaporating drops.
\newblock {\em Langmuir}, 21(18):8226--8233, 2005.

\bibitem{Rastogi:2008ba}
V.~Rastogi, S.~Melle, O.~G. Calder{\'o}n, A.~A. Garc{\'\i}a, M.~M{\'a}rquez,
  and O.~D. Velev.
\newblock {Synthesis of Light-Diffracting Assemblies from Microspheres and
  Nanoparticles in Droplets on a Superhydrophobic Surface}.
\newblock {\em Adv. Mater.}, 20(22):4263--4268, Nov. 2008.

\bibitem{Rio:2006}
E.~Rio, A.~Daerr, F.~Lequeux, and L.~Limat.
\newblock Moving contact lines of a colloidal suspension in the presence of
  drying.
\newblock {\em Langmuir}, 22:3186--3191, 2006.

\bibitem{Ristenpart:2007}
W.~D. Ristenpart, P.~G. Kim, C.~Domingues, J.~Wan, and H.~A. Stone.
\newblock Influence of substrate conductivity on circulation reversal in
  evaporating drops.
\newblock {\em Phys. Rev. Lett.}, 99(23):234502, 2007.

\bibitem{Rossi:2019}
M.~Rossi, A.~Marin, and C.~J. K\"ahler.
\newblock {Interfacial flows in sessile evaporating droplets of mineral water}.
\newblock {\em Phys. Rev. E}, 100(033103):033103, 2019.

\bibitem{Sadek:2015}
C.~Sadek, P.~Schuck, Y.~Fallourd, N.~Pradeau, C.~Le~Floch-Fou\'er\'e, and
  R.~Jeante.
\newblock Drying of a single droplet to investigate
  process–structure–function relationships: a review.
\newblock {\em Dairy Sci. \& Technol.}, 95(771):771--794, 2015.

\bibitem{saenz2015evaporation}
P.~J. S{\'a}enz, K.~Sefiane, J.~Kim, O.~K. Matar, and P.~Valluri.
\newblock Evaporation of sessile drops: a three-dimensional approach.
\newblock {\em J. Fluid Mech.}, 772:705--739, 2015.

\bibitem{Sefiane:2010}
K.~Sefiane.
\newblock On the formation of regular patterns from drying droplets and their
  potential use for biomedical applications.
\newblock {\em J. Bionic Eng.}, 7:S82--S93, 2010.

\bibitem{Sefiane:2014fl}
K.~Sefiane.
\newblock Patterns from drying drops.
\newblock {\em Adv. Colloid Interface Sc.}, 206:372--381, Apr. 2014.

\bibitem{Sefiane:2007he}
K.~Sefiane and C.~A. Ward.
\newblock {Recent advances on thermocapillary flows and interfacial conditions
  during the evaporation of liquids}.
\newblock {\em Adv. Colloid Interface Sci.}, 134-135:201--223, Oct. 2007.

\bibitem{Sempels:1bi}
W.~Sempels, R.~De~Dier, H.~Mizuno, J.~Hofkens, and J.~Vermant.
\newblock {Auto-production of biosurfactants reverses the coffee ring effect in
  a bacterial system}.
\newblock {\em Nat. Commun.}, 4:1757--8, 2013.

\bibitem{Sen:2009}
D.~Sen, S.~Mazumder, J.~Melo, A.~Khan, S.~Bhattyacharya, and S.~D\'Souza.
\newblock Evaporation driven self-assembly of a colloidal dispersion during
  spray drying: Volume fraction dependent morphological transition.
\newblock {\em Langmuir}, 25:6690--6695, 2007.

\bibitem{Sen:2007}
D.~Sen, O.~Spalla, O.~Tach\'e, P.~Haltebourg, and A.~Thill.
\newblock Slow drying of a spray of nanoparticles dispersion. in situ saxs
  investigation.
\newblock {\em Langmuir}, 23:4296--4302, 2007.

\bibitem{seyfert2021evaporation}
C.~Seyfert, E.~J. Berenschot, N.~R. Tas, A.~Susarrey-Arce, and A.~Marin.
\newblock Evaporation-driven colloidal cluster assembly using droplets on
  superhydrophobic fractal-like structures.
\newblock {\em Soft matter}, 17(3):506--515, 2021.

\bibitem{Snoeijer:2013}
J.~H. Snoeijer and B.~Andreotti.
\newblock Moving contact lines: Scales, regimes, and dynamical transitions.
\newblock {\em Annu. Rev. Fluid Mech.}, 45(1):269--292, 2013.

\bibitem{Stauber:2015}
J.~M. Stauber, S.~K. Wilson, B.~R. Duffy, and K.~Sefiane.
\newblock {On the lifetimes of evaporating droplets with related initial and
  receding contact angles}.
\newblock {\em Phys. Fluids}, 27:122101, 2015.

\bibitem{Sternling1959}
C.~V. Sternling and L.~E. Scriven.
\newblock Interfacial turbulence: {H}ydrodynamic instability and the
  {M}arangoni effect.
\newblock {\em AIChE Journal}, 5(4):514--523, 1959.

\bibitem{still2012surfactant}
T.~Still, P.~J. Yunker, and A.~G. Yodh.
\newblock Surfactant-induced marangoni eddies alter the coffee-rings of
  evaporating colloidal drops.
\newblock {\em Langmuir}, 28(11):4984--4988, 2012.

\bibitem{Still:2012vj}
T.~Still, P.~J. Yunker, and A.~G. Yodh.
\newblock {Surfactant-induced Marangoni eddies alter the coffee-rings of
  evaporating colloidal drops}.
\newblock {\em Langmuir}, pages~--, 2012.

\bibitem{Tan:2016kpa}
H.~Tan, C.~Diddens, P.~Lv, J.~G.~M. Kuerten, X.~Zhang, and D.~lohse.
\newblock {Evaporation-triggered microdroplet nucleation and the four life
  phases of an evaporating Ouzo drop}.
\newblock {\em Proc. Natl. Acad. Sci.}, 113(31):8642--8647, Aug. 2016.

\bibitem{Tarasevich:2005}
Y.~Y. Tarasevich.
\newblock Simple analytical model of capillary flow in an evaporating sessile
  drop.
\newblock {\em Phys. Rev. E}, 71(2):027301, 2005.

\bibitem{Lijun2022}
L.~Thayyil~Raju, C.~Diddens, Y.~Li, A.~Marin, M.~van~der Linden, X.~Zhang, and
  D.~Lohse.
\newblock Evaporation of a sessile colloidal water-glycerol droplet: Marangoni
  ring formation.
\newblock {\em TBD}, 2022.

\bibitem{Thiele:2014}
U.~Thiele.
\newblock Patterned deposition at moving contact lines.
\newblock {\em Adv. Colloid Interface Sci.}, 206:399--413, 2014.

\bibitem{thomson1855tears}
J.~Thomson.
\newblock Xlii. on certain curious motions observable at the surfaces of wine
  and other alcoholic liquors.
\newblock {\em Lond. Edinb. Dublin philos. Mag. J. Sci.}, 10(67):330--333,
  1855.

\bibitem{Trantum:2013ff}
J.~R. Trantum, Z.~E. Eagleton, C.~A. Patil, J.~M. Tucker-Schwartz, M.~L.
  Baglia, M.~C. Skala, and F.~R. Haselton.
\newblock {Cross-Sectional Tracking of Particle Motion in Evaporating Drops:
  Flow Fields and Interfacial Accumulation}.
\newblock {\em Langmuir}, 29(21):6221--6231, May 2013.

\bibitem{Tsapis:2002}
N.~Tsapis, D.~Bennett, B.~Jackson, D.~Weitz, and D.~Edwards.
\newblock Trojan particles: Large porous carriers of nanoparticles for drug
  delivery.
\newblock {\em Proc. Natl. Acad. Sci. U.S.A.}, 99(12001):12001--12005, 2002.

\bibitem{Tsapis:2005}
N.~Tsapis, E.~Dufresne, S.~Sinha, C.~Riera, J.~Hutchinson, L.~Mahadevan, and
  D.~Weitz.
\newblock Onset of buckling in drying droplets of colloidal suspensions.
\newblock {\em Phys. Rev. Lett.}, 94(018302):018302, 2005.

\bibitem{van2020marangoni}
R.~van Gaalen, C.~Diddens, H.~Wijshoff, and J.~Kuerten.
\newblock The evaporation of surfactant-laden droplets: A comparison between
  contact line models.
\newblock {\em J. Colloid Interface Sci.}, 579:888--897, 2020.

\bibitem{van2021marangoni}
R.~van Gaalen, C.~Diddens, H.~Wijshoff, and J.~Kuerten.
\newblock Marangoni circulation in evaporating droplets in the presence of
  soluble surfactants.
\newblock {\em J. Colloid Interface Sci.}, 584:622--633, 2021.

\bibitem{van2022marangoni}
R.~van Gaalen, H.~Wijshoff, J.~Kuerten, and C.~Diddens.
\newblock Competition between thermal and surfactant-induced marangoni flow in
  evaporating sessile droplets.
\newblock {\em J. Colloid Interface Sci.}, 622:892--903, 2022.

\bibitem{Isa2021particlesinterfaces}
J.~Vialetto, M.~Zanini, and L.~Isa.
\newblock Attachment and detachment of particles to and from fluid interfaces.
\newblock {\em Cur. Opin. Colloid Interface Sci.}, page 101560, 2021.

\bibitem{practicalguide2020}
W.~Wang, X.~Lv, J.~L. Moran, S.~Duan, and C.~Zhou.
\newblock A practical guide to active colloids: choosing synthetic model
  systems for soft matter physics research.
\newblock {\em Soft Matter}, 16:3846--3868, 2020.

\bibitem{Ward:2004dg}
C.~A. Ward and F.~Duan.
\newblock {Turbulent transition of thermocapillary flow induced by water
  evaporation}.
\newblock {\em Phys. Rev. E}, 69(5):187--10, May 2004.

\bibitem{Weon2013PRE}
B.~Weon and J.~Je.
\newblock Fingering inside the coffee ring.
\newblock {\em Phys. Rev. E}, 87:013003, 2013.

\bibitem{Weon2013selfpinning}
B.~Weon and J.~Je.
\newblock {Self-Pinning by Colloids Confined at a Contact Line}.
\newblock {\em Phys. Rev. Lett.}, 110(2):--, Jan. 2013.

\bibitem{Harting2018friction}
Q.~Xie and J.~Harting.
\newblock From dot to ring: The role of friction in the deposition pattern of a
  drying colloidal suspension droplet.
\newblock {\em Langmuir}, 34(18):5303--5311, 2018.
\newblock PMID: 29652501.

\bibitem{Xu:2007je}
X.~Xu and J.~Luo.
\newblock {Marangoni flow in an evaporating water droplet}.
\newblock {\em Appl. Phys. Lett.}, 91(12):124102, 2007.

\bibitem{Yunker:2011vr}
P.~J. Yunker, T.~Still, M.~A. Lohr, and A.~G. Yodh.
\newblock {Suppression of the coffee-ring effect by shape-dependent capillary
  interactions}.
\newblock {\em Nature}, 476(7360):308--311, 2011.

\bibitem{Zhang:2014}
L.~Zhang, Y.~Nguyen, and W.~Chen.
\newblock ``coffee ring" formation dynamics on molecularly smooth substrates
  with varying receding contact angles.
\newblock {\em Colloids Surf. A: Physicochem. Eng. Aspects}, 449:42--50, 2014.

\end{thebibliography}
\end{document}